\documentclass[journal]{IEEEtran}
\usepackage{graphicx}
\usepackage[cmex10]{amsmath}
\usepackage{cases}
\usepackage{amsthm}
\usepackage{cite}
\usepackage{amssymb}
\usepackage{algorithm}
\usepackage{algorithmic}
\usepackage{multirow}
\usepackage{amsmath}
\usepackage{xcolor}
\usepackage{CJK}
\usepackage{subeqnarray}
\usepackage{cases}
\usepackage{enumerate}
\usepackage{cuted}
\usepackage{booktabs}
\usepackage{array}
\newcolumntype{M}[1]{>{\centering\arraybackslash}m{#1}}
\usepackage{graphicx}
\usepackage{subcaption}
\newtheorem{proposition}{\hskip\parindent\it{Proposition}}
\newtheorem{theorem}{\hskip\parindent \it Theorem}

\ifCLASSINFOpdf
\else
\fi

\begin{document}

\title{Energy-Efficient STAR-RIS Enhanced UAV-Enabled  MEC Networks with Bi-Directional Task Offloading}
\author{Han Xiao,~\IEEEmembership{Student Member,~IEEE,} Xiaoyan Hu$^*$,~\IEEEmembership{Member,~IEEE,}\\
	 Weile Zhang,~\IEEEmembership{Member,~IEEE,} Wenjie Wang,~\IEEEmembership{Member,~IEEE,} \\ %Zhou Su,~\IEEEmembership{Senior Member,~IEEE,}
        %Ang Li,~\IEEEmembership{Member,~IEEE,}
	Kai-Kit~Wong,~\IEEEmembership{Fellow,~IEEE}, Kun~Yang,~\IEEEmembership{Fellow,~IEEE}
	
%\thanks{%This work is supported in part by the National Natural Science Foundation of China (NSFC) under Grant 62201449 and Grant 62071370, in part by the Key R$\&$D Projects of Shaanxi Province under Grant 2023-YBGY-040, in part by the Qin Chuang Yuan High-Level Innovation and Entrepreneurship Talent Program under Grant QCYRCXM-2022-231, and in part by the ``Si Yuan Scholar" Foundation.
%This paper will be presented in part at IEEE Global Communications Conference (GLOBECOM),  Kuala Lumpur, Malaysia, December 2023.
%The associate editor coordinating the review of this article and approving it for publication was Prof. M. Cenk Gursoy.
%\emph{(Corresponding author: Xiaoyan Hu.)}}	
\thanks{%This work is supported in part by the National Natural Science Foundation of China (NSFC) under Grant 62201449 and Grant 62071370, in part by the Young Elite Scientists Sponsorship Program by CAST under Grant No.YESS20230611, in part by the Key R$\&$D Projects of Shaanxi Province under Grant 2023-YBGY-040, in part by the Qin Chuang Yuan High-Level Innovation and Entrepreneurship Talent Program under Grant QCYRCXM-2022-231, and in part by the ``Si Yuan Scholar" Foundation.
	%This paper will be presented in part at IEEE Global Communications Conference (GLOBECOM),  Kuala Lumpur, Malaysia, December 2023.
	%The associate editor coordinating the review of this article and approving it for publication was Prof. M. Cenk Gursoy.
	\emph{(Corresponding author: Xiaoyan Hu.)}}		
	\thanks{H. Xiao, X. Hu, W. Zhang, and W. Wang  are with the School of Information and Communications Engineering, Xi'an Jiaotong University, Xi'an 710049, China. (email: hanxiaonuli@stu.xjtu.edu.cn, xiaoyanhu@xjtu.edu.cn,  \{wlzhang, wjwang\}@mail.xjtu.edu.cn).}%ang.li2020@xjtu.edu.cn, , zhousu@xjtu.edu.cn,
	\thanks{K.-K. Wong is with the Department of Electronic and Electrical Engineering, University College London, London WC1E 7JE, U.K. (email: kai-kit.wong@ucl.ac.uk)}	
	\thanks{K. Yang is with the School of Computer Science and Electronic Engineering, University of Essex, Colchester CO4 3SQ, U.K. (e-mail: kunyang@essex.ac.uk).}
}

\maketitle

\begin{abstract}
This paper introduces a novel multi-user mobile edge computing (MEC) scheme facilitated by the simultaneously transmitting and reflecting reconfigurable intelligent surface (STAR-RIS) and the unmanned aerial vehicle (UAV). Unlike existing MEC approaches, the proposed scheme enables bi-directional offloading, allowing users to concurrently offload tasks to the MEC servers located at the ground base station (BS) and UAV with STAR-RIS support. To evaluate the effectiveness of the proposed MEC scheme, we first formulate an optimization problem aiming at maximizing the energy efficiency of the system while ensuring the quality of service (QoS) constraints by jointly optimizing the resource allocation, user scheduling, passive beamforming of the STAR-RIS, and the UAV trajectory. A block coordinate descent (BCD) iterative algorithm designed with the Dinkelbach's algorithm and the successive convex approximation (SCA) technique is proposed to effectively handle the formulated non-convex optimization problem with significant coupling among variables. Simulation results indicate that the proposed STAR-RIS enhanced UAV-enabled MEC scheme possesses significant advantages in enhancing the system energy efficiency over other baseline schemes including the conventional RIS-aided scheme.
\end{abstract}
\begin{IEEEkeywords}
	STAR-RIS, unmanned aerial vehicle (UAV), mobile edge computing (MEC),  energy efficiency. %, trajectory optimization. %alternating optimization,
\end{IEEEkeywords}
\IEEEpeerreviewmaketitle
\section{Introduction}\label{sec:S1}
Currently, there is an exponential surge in the proliferation of wireless devices, accompanied by a notable diversification in the utilization of Internet of Things (IoT) applications. As for the implementation of those computation-intensive and latency-critical applications, e.g., autonomous driving, augmented and virtual reality, etc.,  it presents significant challenges for the widely used center-cloud computing framework to effectively handle large amounts of data in swift actions \cite{mao2017survey, hu2023irs, al2015internet}. To address this challenge, the technology of mobile edge computing (MEC) has emerged as a promising solution for effectively tackling the significantly augmented computational necessity, since it is able to bring the capabilities of cloud computing to the network edge and enables data processing in close proximity. Consequently the MEC technique can perform well in reducing network congestion, and improving service quality and user experience. The benefits introduced by MEC have garnered significant attention from researchers, and thus many efforts primarily focus on reducing latency, conserving energy, enhancing energy efficiency and so on \cite{chen2014decentralized, chen2021latency, bi2018computation, hu2018wireless, shi2021computation}.
\subsection{Related Works}
While the MEC technology provides an effective means to enhance the network computing capabilities, the traditional placement strategy for MEC servers near the ground base stations (BSs) or access points (APs)  may result in a limited service coverage.
To overcome this limitation, the unmanned aerial vehicles (UAVs) are leveraged to assist the task completion of the MEC networks due to the inherent advantages of UAVs, such as exceptional mobility and flexibility \cite{jeong2017mobile, gu2021uav, xu2021uav, hu2019uav, hu2020Wireless, zhang2019joint, zhang2019computation, yang2021ai, yang2020offloading}. %zhang2020energy,
Specifically,  in \cite{jeong2017mobile}, the UAV is equipped with a MEC server acting as an aerial MEC platform to facilitate the computation of the offloaded tasks for users with low-quality transmission links from the BSs or APs. In \cite{gu2021uav}, the UAV is leveraged as a relay to support the transfer of users' computational tasks to the MEC servers located at BS. The authors in \cite{xu2021uav} introduce a novel MEC scheme that involves both aerial and ground cooperation. The proposed scheme enables users to efficiently offload their task data to multiple base stations (BSs) and UAVs in a collaborative manner.  Furthermore, a two-way offloading UAV-aided MEC scheme is proposed in \cite{hu2019uav, zhang2019joint, hu2020Wireless} to enhance the computing capacity of the MEC network, where the UAV not only carries the MEC processor but also serves as a relay to facilitate the offloading of tasks from users to ground MEC servers. Although UAV is able to effectively improve the computing capacity of MEC networks, the existing UAV-aided MEC schemes are designed by adapting to uncontrollable random wireless channels, which seriously limits the task offloading efficiency.

The reconfigurable intelligent surface (RIS) technique emerges as a promising solution to address this challenge \cite{huang2019reconfigurable, wu2019intelligent, yang2023reconfigurable}. Due to the fact that RIS can dynamically adjust the phases and amplitudes of incident signals, allowing the creation of controllable end-to-end virtual channels, RIS has been incorporated into various kinds of communication systems including MEC networks \cite{hu2021reconfigurable, li2021energy, he2023joint, qin2023joint, xu2022computation, zhai2022energy, duo2023joint, cao2021reconfigurable}.
It is important to highlight that UAVs flying at high altitudes enables the establishment of a reliable line-of-sight (LoS) connection between the UAV and users with a high possibility.
Additionally, RIS technology has the capability to reconfigure the wireless propagation environment, and thus combining UAV and RIS will be a win-win strategy for MEC networks. In particular, a RIS-assisted UAV-enabled MEC scheme considering the aim of maximizing energy efficiency is proposed in \cite{qin2023joint}. This MEC scheme utilizes the RIS positioned on the building to enhance the signals of users' task offloading and direct them towards the MEC server located on the UAV.  To further improve the computational capacity of the MEC network, a two-way offloading scheme assisted by the UAV and the RIS is proposed in \cite{xu2022computation}. In this scheme, a multi-antenna UAV is responsible for processing a portion of users' computation tasks while also acts as a relay to transmit the remaining computation tasks to the BS with the assistance of the RIS installed on the building's surface.

It is worth nothing that the flexibility and mobility of the RIS in traditional RIS-assisted UAV-enabled MEC schemes are limited \cite{qin2023joint,xu2022computation}, considering that the location of the RIS is usually fixed. In order to improve the flexibility and mobility of the RIS, an aerial RIS-aided MEC scheme is proposed to assist the MEC network in \cite{zhai2022energy}. In this scheme, users only offload their tasks to ground MEC server with the help of the reflected ability of the RIS. Actually, the utilization of computing resources on the UAV is significantly limited by the reflection-only RIS, as it redirects all signals intended for task offloading to the BS. A MEC scheme utilizing two UAVs has been introduced by Duo \textit{et al.} in \cite{duo2023joint} to overcome this constraint. In this scheme, one UAV is equipped with a RIS to support and optimize the transmission of task offloading signals from users to the MEC server positioned on the second UAV. It is important to highlight that the UAV housing the MEC server is capable to handle some of the computational tasks and serve as a relay to forward the remaining tasks to the ground BS, thereby significantly enhancing the computational capacity of the MEC network.

\begin{figure}[ht]
	\centering
	\includegraphics[scale=0.4]{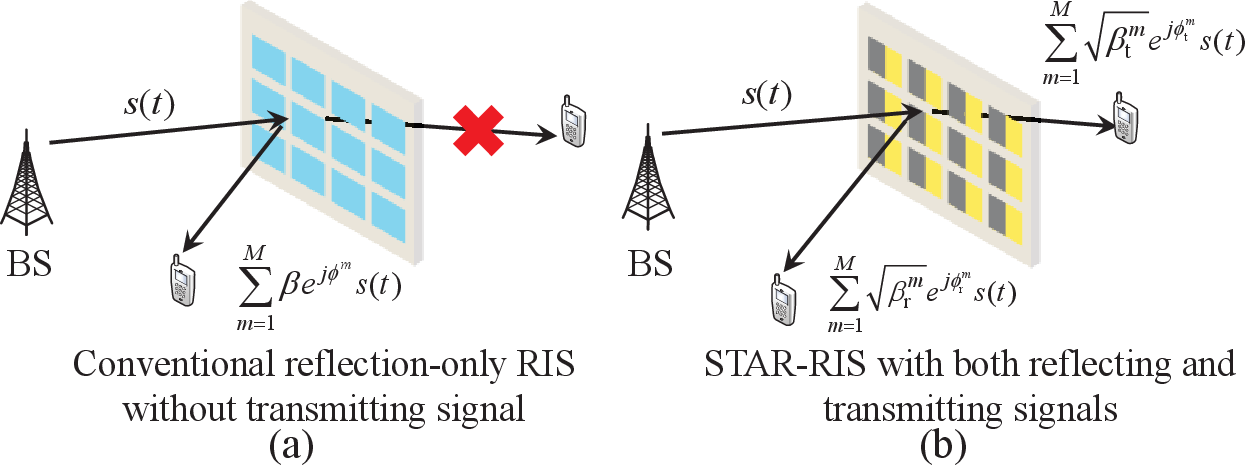}\\
\caption{The main difference between the STAR-RIS and the conventional RIS in signal processing.}\label{fig:RIS_STAR-RIS}
\end{figure}

\textcolor{blue}{Actually, in the existing RIS-aided wireless communication schemes, the traditional RIS is solely able to execute the reflection modulation to the incident signals, which requires the transceiver terminal equipment to be located in the same side of the RIS. In other words, the conventional RIS can only  reconfigure the half-space wireless propagation environment, which will significantly limit the coverage  area of wireless networks and the flexibility in deploying the RIS. To breakthrough this limitation, an advanced RIS technology, named as  simultaneously transmitting and reflecting reconfigurable intelligent surface (STAR-RIS), has drawn great attention from both academia and industry\cite{mu2021simultaneously, zhang2021intelligent, liu2021star,  ahmed2023survey}. STAR-RIS can split the incident signal into two parts, where one part is reflected to the same side of the incident signal and the other part is transited to the opposite side, allowing a 360$^\circ$ coverage, compared with the conventional RIS. As shown in Fig. \ref{fig:RIS_STAR-RIS}, we present the main difference between the STAR-RIS and RIS in signal processing. Specifically, when the BS transmits signals to the traditional RIS, the signals can only be reflected towards the user positioned on the same side as the BS. The user on the opposite side will not receive any signals from the RIS. However, the STAR-RIS has the capability to not only reflect signals to users positioned on the same side as the BS but also transmit signals to users on the opposing side.
}
\subsection{Motivation and Contributions}
\textcolor{blue}{Due to the inherent advantages, STAR-RIS possesses an enormous application potentials in various wireless communication systems, e.g., secure communications \cite{Xiao23STAR-RIS,xiao2023star, xiao2024star}, integrated sensing and communications \cite{liu2023toward, xue2024noma} and MEC networks \cite{qin2023joint_,zhang2023resource, aung2024aerial}.
Specifically, a novel STAR-RIS-assisted MEC scheme is proposed in \cite{aung2024aerial}, where the STAR-RIS is attached vertically on the UAV. With the assistance of STAR-RIS, the ground users distributed in a 360$^\circ$ manner can efficiently offload their tasks to the BS. It is worth noting that the traditional RIS mounted horizontally on the UAV can offer a 360$^\circ$ coverage for ground users, as demonstrated in \cite{zhai2022energy}. However, the aerial STAR-RIS-aided MEC system provides greater flexibility in modulating incident signals through the use of transmission and reflection beamforming. Note that, the MEC scheme supported by the STAR-RIS in \cite{aung2024aerial} still exhibits certain constraints:  (\romannumeral 1) The vertically positioned STAR-RIS will experience air resistance when the UAV is swiftly moving in the air. And the resistance becomes more pronounced as the surface area of the RIS increases, which will affect the flexibility of the UAV.  (\romannumeral 2) The MEC scheme underutilizes the complete spatial modulation potential inherent in the STAR-RIS and it is difficult for this scheme to use the computing resources on the UAV.}

Although the existing RIS-assisted UAV-enabled MEC schemes with the two-way task offloading can fully leverage the computing resources situated at the BS and UAV, they have the following deficiencies: (\romannumeral 1) The UAV serves as multiple roles where it acts as an MEC platform for processing partial offloaded tasks, and a relay to send the remaining tasks to the BS, which imposes a challenge on UAV hardware design.
(\romannumeral 2) These two-step MEC schemes are not energy and time efficient since the UAV needs to receive and decode all the offloaded tasks, and then transmit the unprocessed tasks to the BS in the next time slot.
\textcolor{blue}{The primary driving force behind this study stems from the imperative need to tackle the aforementioned challenges. To overcome limitations of the current MEC schemes, we propose the STAR-RIS enhanced UAV-assisted MEC scheme. In this scheme, the STAR-RIS is attached on the UAV parallel to the ground, which allows users to offload their computing tasks to MEC servers located at the BS and UAV simultaneously,
facilitated by the reflection and transmission features of the
STAR-RIS.}
In this paper, our main contributions are summarized as follows:
\begin{itemize}
	\item \textbf{\textit{STAR-RIS enhanced UAV-enabled MEC Architecture with Bi-directional Offloading:}}  A novel MEC scheme aided by the STAR-RIS horizontally mounted on the UAV is proposed for the first time. In contrast to the existing MEC schemes, the proposed scheme allows users to simultaneously offload their computing tasks to the MEC servers located at the BS and UAV in a bi-directional manner through the reflection and transmission capabilities of the STAR-RIS. Note that the UAV in the proposed MEC scheme is solely responsible for carrying the MEC server and the STAR-RIS to assist the computation and offloading of users' tasks.
	\item \textbf{\textit{Optimization Problem Formulation Maximizing Energy Efficiency under Practical Constraints:}} To assess the effectiveness of the proposed MEC scheme, We first formulate an optimization problem with the aim of maximizing the energy efficiency of the system and ensuring users' quality of service (QoS) constraints by jointly designing the resource allocation, users scheduling, passive beamforming and UAV trajectory. Actually, managing this optimization problem can be quite difficult because of the presence of a fractional objective function and the significant couplings among optimization variables.
	\item \textbf{\textit{Iterative Algorithm with Guaranteed Convergence and Substantial Performance Gain:}} To effectively address this non-convex optimization problem, the alternative strategy is leveraged to divide the optimization problem into three subproblems. Then, an iterative algorithm based on the Dinkelbach's algorithm \cite[Chapter 3.2.1]{zappone2015energy} and the successive convex approximation (SCA) technique is proposed to effectively solve these three subproblems. The assured convergence and efficacy of the algorithm under consideration can be validated through an analysis of its convergence curve and a comparative assessment against the semidefinite relaxation (SDR) technique. Moreover, the potential of the STAR-RIS enhanced UAV-enabled MEC scheme are demonstrated by comparing the simulation results with four other baseline schemes.
\end{itemize}

The remainder of this paper is organized as follows. In Section \ref{sec:S2}, the system model of the STAR-RIS enhanced UAV-enabled MEC network is presented, along with the channel models, task offloading and computation models, as well as the energy consumption model of the system. The formulated optimization problem and the designed iterative algorithm are shown in Section \ref{sec:S3}, including the convergence and complexity analysis of the proposed algorithm. The numerical simulation is conducted in Section \ref{sec:S4} to verify the effectiveness of the designed algorithm and the proposed MEC scheme. Finally, the conclusion is made in Section \ref{sec:S5}.

\textit{Notation:} Operator $\circ$ denotes the Hadamard product. $(\cdot)^T$, $(\cdot)^H$ and $(\cdot)^*$ represent transpose, conjugate transpose and conjugate, respectively. $\operatorname{Diag}(\mathbf{a})$ denotes a diagonal matrix with diagonal elements in vector $\mathbf{a}$ while $\operatorname{diag}(\mathbf{A})$ denotes a vector whose elements are composed of the diagonal elements of matrix $\mathbf{A}$. $|\cdot|$, $\|\cdot\|$ indicate the complex modulus and the spectral norm, respectively. $\mathbb{C}^{M\times1}$ stands for the set of $M\times1$ complex vectors. Operator $\operatorname{norm}(\mathbf{a})$ will normalize the amplitude of all entities in vector $\mathbf{a}$ as 1. Operator $\operatorname{arg}(\mathbf{a})$ denotes the operation of extracting the phase angle of the complex number.
\section{System Model}\label{sec:S2}
\begin{figure}[ht]
	\centering
	\includegraphics[scale=0.4]{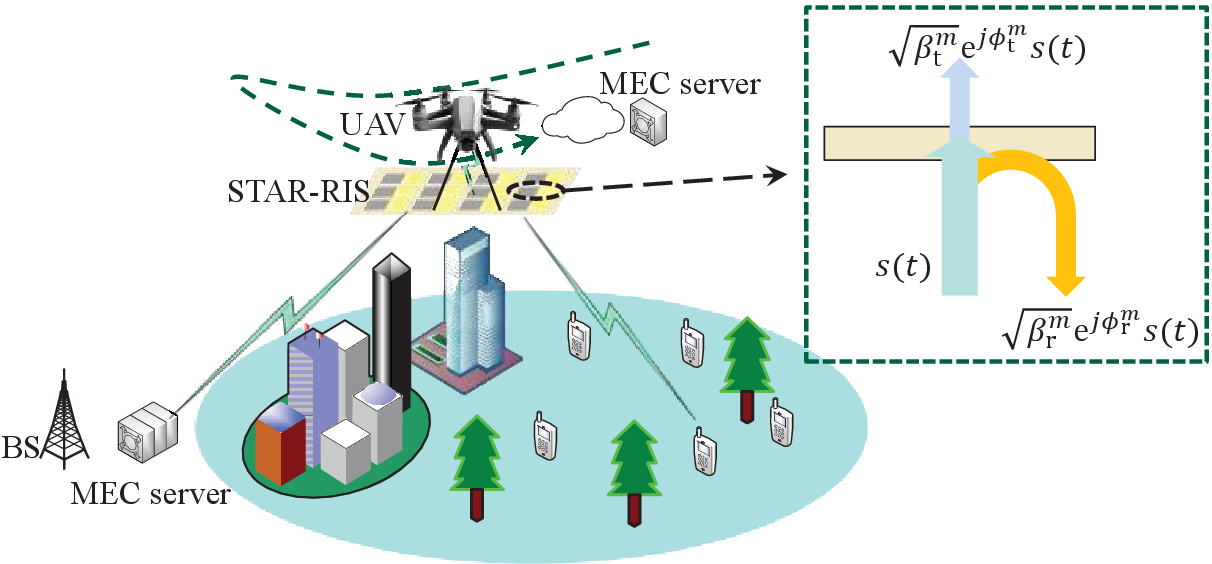}\\
\caption{The UAV-enabled MEC network with bi-directional offloading strategy supported by the STAR-RIS.}\label{fig:System model}
\end{figure}

Fig. \ref{fig:System model} illustrates the architecture of the UAV-enabled MEC network with the assistance of the STAR-RIS. The network comprises a ground base station (BS), $K$ users each equipped with a single antenna, a UAV equipped with a signal antenna and installed with a STAR-RIS featuring $M$ elements, along with two MEC servers situated respectively at the BS and the UAV. This study adopts the energy splitting protocol for the STAR-RIS, wherein all elements incorporated within the STAR-RIS possess the capability to simultaneously reflect (R) and transmit (T) incident signals \cite{mu2021simultaneously}. \textcolor{blue}{Specifically, when signals from users are incident on STAR-RIS, a portion of the signals are reflected to the BS by STAR-RIS, while the remaining portion of the signals reach the UAV via the STAR-RIS.} This feature enables users to offload their computing tasks in a bidirectional manner  concurrently to the MEC servers located at the BS and the UAV, respectively.

In this paper, we divide  the mission period $T$ into $N$ equal time slots, i.e., $\delta_\mathrm{t}=T/N$, which is sufficiently small. Considering that the UAV is equipped with a transceiver that has a single antenna, the time division multiple access (TDMA) protocol is implemented to handle users' offloaded tasks, requiring that only one user will be chosen to offload the computing tasks to the BS and the UAV within a time slot. Here, we use the variable $\zeta_k[n]\in\{0,1\}$ for $k\in\mathcal{K}\triangleq\{1, \cdots, K\}, n\in\mathcal{N}\triangleq\{1, \cdots, N\}$  to represent the user association decision for task offloading in a time slot. In particular, if $\zeta_k[n]=1$, it means that the $k$-th user  is chosen to offload its task  to the BS and UAV in the $n$-th time slot with the assistance of the STAR-RIS. \textcolor{blue}{To ensure that only one user is selected to offload its tasks in each time slot, variable $\zeta_k[n]$ needs to satisfy the following constraint:}
\begin{align}
\sum\limits_{k=1}^K\zeta_k[n]=1,~\zeta_k[n]\in\{0,1\},~\forall k \in \mathcal{K},~ n \in\mathcal{N}.
\end{align}

In order to clearly describe the considered scenario, we assume that all nodes are situated in a 3D Cartesian coordinate system.
The positions of the BS and the $k$-th user are respectively denoted as $\mathbf{q}_{\mathrm{BS}}=[x_{\mathrm{BS}}, y_{\mathrm{BS}}, z_{\mathrm{BS}}]^T$ and  $\mathbf{q}_{k}=[x_{k}, y_{k}, 0]^T$, where $x_\mathrm{BS}$ and $x_{k}$ respectively denote the abscissa value of the BS and the $k$-th user, $y_\mathrm{BS}$ and $y_{k}$ respectively represent the vertical coordinates of the BS and the $k$-th user, $z_{\mathrm{BS}}$ is the height of the BS. It is assumed that the UAV flies at a fixed altitude $H$ and its position remains constant within a given time slot considering the small value of $\delta_\mathrm{t}$. Consequently, the location of the UAV in the $n$-th time slot can be represented as $\mathbf{q}_{\mathrm{ua}}[n]=[x_{\mathrm{ua}}[n], y_{\mathrm{ua}}[n], H]^T$, $n\in\mathcal{N}$, which should adhere to the subsequent flight constraints:
\begin{align}
&	\mathbf{v}_\mathrm{ua}[n]=\frac{\mathbf{q}_\mathrm{ua}[n+1]-\mathbf{q}_\mathrm{ua}[n]}{\delta _\mathrm{t}},~ \left\|\mathbf{v}_\mathrm{ua}[n]\right\|\leq v_{\max},\\
&\mathbf{a}_\mathrm{ua}[n]=\frac{\mathbf{v}_\mathrm{ua}[n+1]-\mathbf{v}_\mathrm{ua}[n]}{\delta _\mathrm{t}},~ \left\|\mathbf{a}_\mathrm{ua}[n]\right\|\leq a_{\max},\\
&\mathbf{q}_\mathrm{ua}[0]=\mathbf{q}_\mathrm{I},~ \mathbf{q}_\mathrm{ua}[N+1]=\mathbf{q}_\mathrm{F},
\end{align}
where $\left\|\mathbf{v}_\mathrm{ua}[n]\right\|$ and $\left\|\mathbf{a}_\mathrm{ua}[n]\right\|$ respectively represent the flight speed and the acceleration of the UAV in the $n$-th time slot, with the maximum values of  $v_{\max}$ and $a_{\max}$.
%are the maximum flight velocity and acceleration of the UAV, respectively. $\mathbf{q}_\mathrm{I}$ and $\mathbf{q}_\mathrm{F}$ represent the initial point and final point of the UAV trajectory.
In terms of $\mathbf{q}_\mathrm{I}$ and $\mathbf{q}_\mathrm{F}$, they can serve as the ground station where the UAV can access a reliable power supply and receive necessary maintenance.
\subsection{Channel Model}
Due to the fact that the UAV flights at a high altitude, we assume that the line-of-sight (LoS) channels between the ground users/BS and aerial UAV/STAR-RIS can always be guaranteed. It is assumed that the STAR-RIS adopts the uniform planar array (UPA) with $M_x$ elements along $x$-axis direction and $M_y$ elements along $y$-axis direction, i.e., $M=M_xM_y$. Hence, the channel between the $k$-th user ($\varsigma=\mathrm{r}k$)/BS ($\varsigma=\mathrm{rb}$) and the STAR-RIS in the $n$-th time slot for $k\in\mathcal{K},~n\in\mathcal{N}$ can be expressed as \cite{li2021robust}
\begin{align}
	\mathbf{h}_{\varsigma}[n]=\sqrt{\frac{\rho }{d_{\varsigma}^{\alpha_\varsigma}[n]}}\widehat{\mathbf{h}}_{\varsigma}[n]\in\mathbb{C}^{M\times 1},~\varsigma\in\{\mathrm{r}k, \mathrm{rb}\},%~k\in\mathcal{K},~n\in\mathcal{N}
\end{align}
where
\begin{align}
	& \widehat{\mathbf{h}}_{\varsigma}[n]=\notag\\
	&\big[1, \cdots, e^{-j\frac{2\pi d}{\lambda}\left(m_\mathrm{x}-1\right)\xi_{\varsigma}[n]},\cdots, e^{-j\frac{2\pi d}{\lambda}\left(M_\mathrm{x}-1\right)\xi_{\varsigma}[n]}\big]^T
	\otimes\notag\\& \big[1,\cdots, e^{-j\frac{2\pi d}{\lambda}\left(m_\mathrm{y}-1\right)\chi_{\varsigma}[n]},\cdots, e^{-j\frac{2\pi d}{\lambda}\left(M_\mathrm{y}-1\right)\chi_{\varsigma}[n]}\big]^T,
\end{align}
with $\rho=(\frac{\lambda}{4\pi})^2$  \cite{mao2023energy} represents the path loss at a reference distance of $1$ meter (m) with $\lambda$ being the wavelength of the carrier frequency, $\alpha_{\varsigma}$ indicating the pass loss exponent, $d_{\varsigma}[n]$ denoting the distance between $k$-th user/BS and the STAR-RIS. $d$ denotes the adjacent element separation of the STAR-RIS. In addition, the $\xi_{\varsigma}[n]$ and $\chi_{\varsigma}[n]$ corresponding to the $k$-th user and the BS are respectively calculated as
\begin{align}
	&\xi_{\mathrm{r}k}[n]=\cos(\phi_{\mathrm{r}k}[n])\sin(\theta_{\mathrm{r}k}[n])=\frac{x_\mathrm{ua}[n]-x_k}{\left\|\mathbf{q}_\mathrm{ua}[n]-\mathbf{q}_{k}\right\|},\\	&\xi_\mathrm{rb}[n]=\cos(\phi_\mathrm{rb}[n])\sin(\theta_\mathrm{rb}[n])=\frac{x_\mathrm{BS}-x_\mathrm{ua}[n]}{\left\|\mathbf{q}_\mathrm{BS}-\mathbf{q}_\mathrm{ua}[n]\right\|},\\
&\chi_{\mathrm{r}k}[n]=\sin(\phi_{\mathrm{r}k}[n])\sin(\theta_{\mathrm{r}k}[n])=\frac{y_\mathrm{ua}[n]-y_k}{\left\|\mathbf{q}_\mathrm{ua}[n]-\mathbf{q}_{k}\right\|},\\
& \chi_\mathrm{rb}[n]=\sin(\phi_\mathrm{rb}[n])\sin(\theta_\mathrm{rb}[n])=\frac{y_\mathrm{BS}-y_\mathrm{ua}[n]}{\left\|\mathbf{q}_\mathrm{BS}-\mathbf{q}_\mathrm{ua}[n]\right\|}.
\end{align}

It is important to note that the connection between the UAV and STAR-RIS should be described as the near-field channel, denoted as $\mathbf{h}_\mathrm{ru}$, considering the fact that  the distance between the UAV and STAR-RIS is extremely small. Thus, $\mathbf{h}_\mathrm{ru}$ can be expressed as \cite{zhang2022beam} \begin{align}
	\mathbf{h}_\mathrm{ru}=\boldsymbol{\alpha}\circ\mathbf{a}_\mathrm{ru}\in\mathbb{C}^{M\times 1},
\end{align}
where
\begin{itemize}
	\item$\boldsymbol{\alpha}=\Big[\frac{\lambda}{4\pi r_1}, \cdots, \frac{\lambda}{4\pi r_m}, \cdots, \frac{\lambda}{4\pi r_M}\Big]^T,$
\vspace{2mm}
	\item$\mathbf{a}_\mathrm{ru}=\Big[e^{-j\frac{2\pi r_1}{\lambda}}, \cdots, e^{-j\frac{2\pi r_m}{\lambda}}, \cdots, e^{-j\frac{2\pi r_M}{\lambda}}\Big]^T,$
\vspace{2mm}
\end{itemize}
with $r_m$ denotes the distance between the antenna located on the UAV and the $m$-th element of the STAR-RIS, where $m\in\mathcal{M}\triangleq\{1, \cdots, M\}$. Notably, the channel $\mathbf{h}_\mathrm{ru}$ remains invariant throughout, attributed to the fixed spatial relationship maintained between the UAV and the STAR-RIS.
\subsection{Task Offloading and Computation Model}
The offloading rates achieved by the $k$-th user in the $n$-th time slot to the BS and UAV   are respectively given by %for $k\in\mathcal{K},~n\in\mathcal{N}$
\begin{align}
&R^\mathrm{ua}_{k}[n]= \zeta_k[n]B\log_2\Big(1+\frac{p\left|\mathbf{h}^H_\mathrm{ru}\boldsymbol{\Theta}_\mathrm{t}^H[n]\mathbf{h}_{\mathrm{r}k}[n]\right|^2}{\sigma_\mathrm{ua}^2}\Big),\\	
&R^\mathrm{BS}_k[n]=\zeta_k[n]B\log_2\Big(1+\frac{p\left|\mathbf{h}_\mathrm{rb}^H[n]
    \boldsymbol{\Theta}_\mathrm{r}^H[n]\mathbf{h}_{\mathrm{r}k}[n]\right|^2}{\sigma^2_\mathrm{BS}}\Big),
\end{align}
where $p$ denotes the unified transmitted power of users, $B$ is the bandwidth of the system, $\sigma^2_\mathrm{ua}$ and $\sigma^2_\mathrm{BS}$ respectively denote the noise power at the BS and the UAV.
In addition,  $\boldsymbol{\Theta}_\kappa[n]=\operatorname{Diag}\big\{\beta^1_\kappa[n]e^{j\phi^1_{\kappa}[n]}, \cdots, \beta^m_\kappa[n]e^{j\phi^m_{\kappa}[n]}, \cdots, \beta^M_\kappa[n]e^{j\phi^M_{\kappa}[n]} \big\}$ is the matrix of the STAR-RIS's coefficients with $\kappa\in\{\mathrm{r}, \mathrm{t}\}$ representing the reflected or transmitted coefficients of the STAR-RIS, where the amplitudes $\beta^m_\kappa[n]$ and phases $\phi^m_{\kappa}[n]$ of the STAR-RIS should satisfy: $\beta^m_\mathrm{r}[n],~ \beta^m_\mathrm{t}[n]\in(0, 1]$ $(\beta^m_\mathrm{r}[n])^2+(\beta^m_\mathrm{t}[n])^2=1$, $\phi^m_{\mathrm{r}}[n], \phi^m_{\mathrm{t}}[n]\in[0, 2\pi], ~\forall m\in\mathcal{M}$.

It is assumed that the ground users are with limited resources for local computing and thus users' tasks have to be offloaded to the BS and UAV for computing\footnote{\textcolor{blue}{The local processing at the users is disregarded due to the following reasons: (\romannumeral 1) For certain fundamental devices such as smart sensors, edge cameras, surveillance systems, and health monitoring devices, carrying out local computing tasks may pose various challenges.
%		 These devices are typically designed to be compact with limited hardware resources, especially in terms of computational power and memory.
 (\romannumeral 2) In situations where power resources are severely limited and not promptly replenished, devices with limited computing capabilities may opt to suspend local processing in order to conserve power. This is done with the aim of maximizing the device's operational time and ensuring the continuous functioning of essential features.}}.
Let $l^\mathrm{BS}_k[n]$ and $l^\mathrm{ua}_k[n]$ denote the number of the offloaded bits that need to be computed at the BS and UAV for user $k$ in the $n$-th time slot, respectively. Due to the fact that the BS and UAV can only deal with the tasks they have received, the following constraints should be satisfied
\begin{align}
	&	\delta_\mathrm{t}\zeta_k[n]R^\mathrm{ua}_{k}[n]\geq l^\mathrm{ua}_k[n],~ \forall k\in\mathcal{K},~ n
	\in\mathcal{N}, \label{eq_off_ua}\\
	&\delta_\mathrm{t}\zeta_k[n]R^\mathrm{BS}_{k}[n]\geq l^\mathrm{BS}_k[n],~ \forall k\in\mathcal{K},~ n
	\in\mathcal{N}.\label{eq_off_BS}
\end{align}

In order to ensure that every user's minimum computational need is met, we implement the following QoS constraints
\begin{align}
	\sum_{n=1}^N\left(l^\mathrm{BS}_k[n]+l^\mathrm{ua}_k[n]\right)\geq L_k,~ \forall k\in\mathcal{K}, \label{eq_QoS}
\end{align}
where $L_k$ is the minimum computing task requirement of the $k$-th user.

 \textcolor{blue}{Let $f_k^\mathrm{BS}[n]$ and $f_k^\mathrm{ua}[n]$ respectively denote the computation frequency allocated by the BS and UAV for the $k$-th user at the $n$-th time slot.
 In addition, considering the processing causality, it is assumed that the BS and the UAV solely receive the offloaded tasks without carrying out any computations within the first time slot, and the users cease the act of offloading their tasks at the last time slot, i.e.,  $f^\mathrm{BS}_k[1]=f^\mathrm{ua}_k[1]=l^\mathrm{BS}_k[N]=l^\mathrm{ua}_k[N]=0$ for $k\in\mathcal{K}$. In order to guarantee that all users' offloaded task-input data can be completely computed within the give mission time $T$, we have the following information-causality constraints:
\begin{align}
 &\sum\limits_{k=1}^Kf^\mathrm{BS}_k[n]\leq F_\mathrm{BS},~\sum\limits_{k=1}^Kf^\mathrm{ua}_k[n]\leq F_\mathrm{ua}, ~\forall n\in\mathcal{N},\label{eq_freqency_const}	\\ &\sum\limits_{i=2}^{n}\frac{f^\mathrm{ua}_k[i]\delta_\mathrm{t}}{\varrho_\mathrm{ua}}\geq\sum\limits_{i=1}^{n-1} l^\mathrm{ua}_k[i],~ \forall k\in\mathcal{K},~ n\in\mathcal{N}_1, \label{eq_information_causality_const_ua}\\
&\sum\limits_{i=2}^{n}\frac{f^\mathrm{BS}_k[i]\delta_\mathrm{t}}{\varrho_\mathrm{BS}}\geq\sum\limits_{i=1}^{n-1} l^\mathrm{BS}_k[i], ~ \forall k\in\mathcal{K},~  n\in\mathcal{N}_1,\label{eq_information_causality_const_BS}
\end{align}
where $F_\mathrm{BS}$ and $F_\mathrm{ua}$ are the maximum CPU frequency at the BS and the UAV, respectively.} $\varrho_\mathrm{BS}$ and $\varrho_\mathrm{ua}$ represent the number of CPU cycles needed for processing 1-bit of task-input data at the BS and UAV,  respectively, and $\mathcal{N}_1\triangleq\{2, 3, \cdots, N\}$ is a subset of $\mathcal{N}$.
Therefore, the total amount of the completed task-input data for all users within the whole period can be calculated as
\begin{align}
L_\mathrm{tol}=\sum_{n=1}^N\sum_{k=1}^Kl^\mathrm{BS}_k[n]+l^\mathrm{ua}_k[n],
\end{align}
which is an important indicator to measure the computing capability of the MEC system.

\subsection{Energy Consumption Model}
The energy consumption in the system primarily occurs in three ways: task offloading by users, task computation by the BS and UAV, and the UAV's flying process. Specifically, the energy consumed by users to offload tasks during the total mission period is given by
\begin{align}
	E_\mathrm{ut}= p\delta_\mathrm{t}(N-1),
\end{align}
considering the fact that only the  first $N-1$ time slots are utilized by users for task offloading.

According to \cite{hu2019uav}, the energy consumption of the MEC servers situated at the BS and the UAV for computing the received tasks within the whole mission period $T$ can be respectively expressed as
\begin{align}
&E^\mathrm{com}_\mathrm{BS}=\sum_{n=1}^N\sum_{k=1}^K\iota_\mathrm{BS}\delta_\mathrm{t}(f^\mathrm{BS}_k[n])^3,\\
&E^\mathrm{com}_\mathrm{ua}=\sum_{n=1}^N\sum_{k=1}^K\iota_\mathrm{ua}\delta_\mathrm{t}(f^\mathrm{ua}_k[n])^3, \label{eq_ua_energy}
\end{align} 	
where $\iota_\mathrm{BS}$ and $\iota_\mathrm{ua}$  are the effective capacitance coefficients of the MEC servers at the BS and the UAV, respectively.

In this paper, we select the rotary-wing UAV to enhance the STAR-RIS-assisted UAV-enabled MEC network. Consequently, the flying energy consumed by the UAV during the mission period can be expressed as \cite{zeng2019energy}
\begin{align}
	&E^\mathrm{fly}_\mathrm{ua}=\sum_{n=1}^N\delta_\mathrm{t}\Big(P_0\big(1+\frac{3v^2[n]}{U_\mathrm{tip}^2}\Big)+\frac{1}{2}\mu\psi qAv^3[n]\notag\\
	&+\widehat{P}_\mathrm{0}\sqrt{\Big(1+\frac{v^4[n]}{4v_0^4}\Big)^{\frac{1}{2}}-\frac{v^2[n]}{2v_0^2}}\Big),
\end{align}
where $P_0$ and $\widehat{P}_0$ denote the blade profile power and induced power in the hovering status, respectively. $U_\mathrm{tip}$, $\mu$, $\psi$, $q$, $A$ and $v_0$ are parameters related to the UAV's aerodynamics, and more details are presented in Table I of \cite{zeng2019energy}. It is worth noting that $v[n]=\left\|\mathbf{v}_\mathrm{ua}[n]\right\|$ represents the flight velocity of the UAV at the $n$-th time slot.

In this paper, the energy consumption of the BS in processing an computing the received tasks is not taken into account for optimization and analysis, due to the fact that the BS is usually adequately supplied with grid power. Hence, the system's energy consumption is represented by the total energy used by the UAV and all users, given as
\begin{align}
	E_\mathrm{tol}=E_\mathrm{ut}+E_\mathrm{ua}^\mathrm{com}+E^\mathrm{fly}_\mathrm{ua},
\end{align}
which is also an important performance indicator to measure the MEC system.

\section{Problem Formulation and Algorithm Design}\label{sec:S3}
\subsection{Optimization Problem Formulation}
In this section, the optimization problem will be formulated based on the analysis in Section \ref{sec:S2}. In particular, we try to maximize the energy efficiency of the MEC system, defined as $\frac{L_\mathrm{tol}}{E_\mathrm{tol}}$, which takes both the indicators of computing capability $L_\mathrm{tol}$ and the energy consumption $E_\mathrm{tol}$ into consideration, while ensuring the QoS constraints of users with minimum computational requirements, by jointly optimizing
%the resource allocation variable $\mathbf{L}\triangleq\{ l^\mathrm{ua}_k[n], l^\mathrm{BS}_k[n], f^\mathrm{ua}_k[n], f^\mathrm{BS}_k[n], k\in\mathcal{K}, n\in\mathcal{N}\}$, user scheduling variable $\boldsymbol{\zeta}\triangleq\{\zeta_k[n], k\in\mathcal{K}, n\in\mathcal{N}\}$, passive beamforming variable $\boldsymbol{\Upsilon}\triangleq\{\boldsymbol{\Theta}_\mathrm{r}[n], \boldsymbol{\Theta}_\mathrm{t}[n],  n\in\mathcal{N}\}$ and UAV trajectory variable $\mathbf{Q}\triangleq\{\mathbf{q}_\mathrm{ua}[n], n\in\mathcal{N}\}$.
\begin{itemize}
  \item the resource allocation variables in $\mathbf{L}\triangleq\{ l^\mathrm{ua}_k[n], l^\mathrm{BS}_k[n]$, $f^\mathrm{ua}_k[n], f^\mathrm{BS}_k[n],~ k\in\mathcal{K},~ n\in\mathcal{N}\}$;
  \vspace{1mm}
  \item the user scheduling variables for task offloading in $\boldsymbol{\zeta}\triangleq\{\zeta_k[n],~ k\in\mathcal{K}, ~n\in\mathcal{N}\}$;
  \vspace{1mm}
  \item the passive beamforming variables of STAR-RIS in $\boldsymbol{\Upsilon}\triangleq\{\boldsymbol{\Theta}_\mathrm{r}[n]$, $\boldsymbol{\Theta}_\mathrm{t}[n], ~ n\in\mathcal{N}\}$;
  \vspace{1mm}
  \item the UAV trajectory variables in $\mathbf{Q}\triangleq\{\mathbf{q}_\mathrm{ua}[n],~ n\in\mathcal{N}\}$.
  \vspace{1mm}
\end{itemize}

Hence, the corresponding optimization problem can be formulated as
\begin{subequations}\label{eq_orig_opti}
	\begin{align}
		&\max _{\mathbf{L}, \boldsymbol{\zeta}, \boldsymbol{\Upsilon}, \mathbf{Q}}~~ \frac{L_\mathrm{tol}}{E_\mathrm{tol}},\notag \\
		&\quad\text { s.t. }~~\eqref{eq_off_ua}-\eqref{eq_information_causality_const_BS},\label{eq_orig_opti_1}\\
	&\quad\left\|\mathbf{v}_\mathrm{ua}[n]\right\|\leq v_{\max}, \left\|\mathbf{a}_\mathrm{ua}[n]\right\|\leq a_{\max}, ~ \forall n\in\mathcal{N},\label{eq_orig_opti_2}\\
		&\quad\mathbf{q}_\mathrm{ua}[0]=\mathbf{q}_\mathrm{I}, \mathbf{q}_\mathrm{ua}[N]=\mathbf{q}_\mathrm{F},\label{eq_orig_opti_3}\\
		&\quad\sum\limits_{k=1}^K\zeta_k[n]=1, \zeta_k[n]\in\{0,1\},~ n\in\mathcal{N}, k\in\mathcal{K}, \label{eq_orig_opti_4}\\
		&\quad f^\mathrm{ua}_k[1]=f^\mathrm{BS}_k[1]= l^\mathrm{ua}_k[N]=l^\mathrm{BS}_k[N]=0,~ \forall k\in\mathcal{K}, \label{eq_orig_opti_5}\\	
		&\quad (\beta_\mathrm{r}^m[n])^2+(\beta_\mathrm{t}^m[n])^2=1,~  \forall m\in\mathcal{M}, ~n\in\mathcal{N},\label{eq_orig_opti_6}\\
		&\quad\beta_\mathrm{r}^m[n], \beta_\mathrm{t}^m[n] \in(0,1],~ \forall m\in\mathcal{M},~ n\in\mathcal{N}, \label{eq_orig_opti_7}\\
		&\quad\phi_\mathrm{r}^m[n], \phi_\mathrm{t}^m[n]\in[0, 2\pi),~  \forall m\in\mathcal{M},~ n\in\mathcal{N}. \label{eq_orig_opti_8}
	\end{align}
\end{subequations}

Actually, the problem \eqref{eq_orig_opti} is a non-convex problem due to the non-convexity of the fractional objective function, and constraints \eqref{eq_off_ua},  \eqref{eq_off_BS} and \eqref{eq_orig_opti_6}, which is difficult to solve directly. \textcolor{blue}{To effectively handle the optimization problem with the fractional objective function, Dinkelbach's algorithm, as one of the most popular fractional programming algorithm, will be leveraged. According to its core principle, we first transform the problem \eqref{eq_orig_opti} as
\begin{subequations}\label{eq_orig_opti_td}
	\begin{align}
		&\max _{\boldsymbol{\Gamma},\psi}~~ L_\mathrm{tol}-\psi E_\mathrm{tol},\notag \\
		&\quad\text { s.t. }~~\eqref{eq_orig_opti_1}-\eqref{eq_orig_opti_8},\label{eq_orig_opti_td_1}
	\end{align}
\end{subequations}
where $\psi$ denotes the introduced auxiliary variable, $\boldsymbol{\Gamma}=\{\mathbf{L}, \boldsymbol{\zeta}, \boldsymbol{\Upsilon}, \mathbf{Q}\}$. If $\psi^\mathrm{opt}=\frac{L_\mathrm{tol}(\boldsymbol{\Gamma}^\mathrm{opt})}{E_\mathrm{tol}(\boldsymbol{\Gamma}^\mathrm{opt})}$ denotes the optimal objective value of the optimization problem \eqref{eq_orig_opti},  $L_\mathrm{tol}(\boldsymbol{\Gamma})-\psi^\mathrm{opt} E_\mathrm{tol}(\boldsymbol{\Gamma})\leq0$ always holds for any $\boldsymbol{\Gamma}$ and the equality occurs exclusively when the optimal solution $\boldsymbol{\Gamma}^\mathrm{opt}$ of problem \eqref{eq_orig_opti} is reached. Therefore, the optimization problems \eqref{eq_orig_opti_td} and \eqref{eq_orig_opti} will converge to the same optimal solution when $\psi=\psi^\mathrm{opt}$. This equivalence allows us to resolve the transformed problem \eqref{eq_orig_opti_td} to attain the optimal solution for the original problem \eqref{eq_orig_opti}. However, it is challenging to obtain $\psi^\mathrm{opt}$ in advance. To address this challenge, the Dinkelbach's algorithm opts to iteratively update the $\psi$ based on the solution of the transformed problem \eqref{eq_orig_opti_td} to gradually achieve the optimal solution of the optimization problem \eqref{eq_orig_opti}. Hence, the Dinkelbach's algorithm demonstrates considerable potential in efficiently tackling optimization problems with fractional objective functions. More details about the Dinkelbach's algorithm are presented in \cite[Chapter 3.2.1]{zappone2015energy}.
}

 On the basis of the analysis above, the original problem \eqref{eq_orig_opti} can be transformed as the following problem in the $(l+1)$-th iteration of the Dinkelbach's algorithm, which is given by
\begin{subequations}\label{eq_orig_opti_dink}
	\begin{align}
		&\max _{\boldsymbol{\Gamma}}~~ L_\mathrm{tol}-\psi^{(l)} E_\mathrm{tol},\notag \\
		&\quad\text { s.t. }~~\eqref{eq_orig_opti_1}-\eqref{eq_orig_opti_8},\label{eq_orig_opti_dink_1}
	\end{align}
\end{subequations}
where $\psi^{(l)}=\frac{ L_\mathrm{tol}^{(l)}}{E_\mathrm{tol}^{(l)}}$.
Note that the optimization problem \eqref{eq_orig_opti_dink} is still a non-convex optimization problem due to the significant coupling among variables in constraints \eqref{eq_off_ua} and \eqref{eq_off_BS}, as well as the equality constraint \eqref{eq_orig_opti_6}.  To overcome this challenge, the alternative strategy is employed to divide the optimization problem \eqref{eq_orig_opti_dink} into three subproblems.  The algorithm is designed by alternatively optimizing three variable subsets, which are respectively denoted as $\boldsymbol{\Xi}_1=\{\mathbf{L}, \boldsymbol{\zeta}\}$, $\boldsymbol{\Xi}_2=\{\mathbf{L}, \boldsymbol{\Upsilon}\}$ and $\boldsymbol{\Xi}_3=\{\mathbf{L}, \mathbf{Q}\}$. More details of the algorithm design is given in the next subsection.
%where the first subproblem focuses on designing the resource allocation and user scheduling, the passive beamformer is optimized by the second subproblem, and the UAV trajectory is designed by handling the last subproblem.
\subsection{Algorithm Design}
\subsubsection{Designing $\boldsymbol{\Xi}_1$ with the given $\mathbf{Q}$ and $\boldsymbol{\Upsilon}$} First, we jointly optimize the resource allocation variable $\mathbf{L}$ and user scheduling variable $\boldsymbol{\zeta}$ with the given passive beamforming and UAV trajectory. In this case, the original problem \eqref{eq_orig_opti} can be simplified as
\begin{subequations}\label{eq_opti_resource_association}
	\begin{align}
		&\max _{\boldsymbol{\Xi}_1}~ L_\mathrm{tol}(\boldsymbol{\Xi}_1)-\psi^{(l)} E_\mathrm{tol}(\boldsymbol{\Xi}_1),\notag\\
		&\quad\text { s.t. }\eqref{eq_orig_opti_1},~\eqref{eq_orig_opti_4},~\eqref{eq_orig_opti_5}.\label{eq_opti_resource_association_1}
 	\end{align}
\end{subequations}
It is important to highlight that the optimization problem \eqref{eq_opti_resource_association} is a non-convex problem because of  the binary variable $\boldsymbol{\zeta}$.
% To tackle this problem, we first leverage the Dinkelbach's algorithm \cite[Chapter 3.2.1]{zappone2015energy} to transform the fractional objective function. Specifically, the objective function in the $(l+1)$-th iteration of the Dinkelbach's algorithm can be expressed as
%\begin{align}
%	\widehat{\gamma}^{(l+1)}=L_\mathrm{tol}(\boldsymbol{\Xi}_1)-\psi^{(l)}E_\mathrm{tol}(\boldsymbol{\Xi}_1),
%\end{align}
%where $\psi^{(l)}$ can be updated with $\psi^{(l)}=\frac{L_\mathrm{tol}(\boldsymbol{\Xi}_1^{(l)})}{E_\mathrm{tol}(\boldsymbol{\Xi}_1^{(l)})}$.
To address this optimization problem, the non-convex binary constraint \eqref{eq_orig_opti_4} is equivalently transformed as the following constraints:
\begin{align}
	&\sum_{k=1}^K\zeta_k[n]=1, 0\leq\zeta_k[n]\leq1, ~\forall n\in\mathcal{N},~ k\in\mathcal{K},\label{eq_users_scheduling}\\
&\eta_k[n]=\zeta_k[n]-\zeta_k^2[n]=0, ~\forall n\in\mathcal{N}, ~k\in\mathcal{K}.
\end{align}
Note that for any $\zeta_k[n]\in[0,1]$, $\eta_k[n]\geq0$ always holds and the equality in \eqref{eq_users_scheduling} is satisfied if and only if $\zeta_k[n]=0$ or $\zeta_k[n]=1$. Considering the non-negative characteristic of $\{\eta_k[n]\}_{k\in\mathcal{K},n\in\mathcal{N}}$, we try to add the sum of them into the objective function as a penalty term that is subtracted from the objective function. To guarantee the fulfilment of the binary constraint \eqref{eq_orig_opti_4}, an additional inner loop iteration is incorporated into the $(l+1)$-th iteration of the Dinkelbach's algorithm to iteratively enforce the penalty term approaching to 0. Note that the inclusion of the penalty term in the objective function results in a non-concave objective function due to the convex nature of $-\eta_k[n]$ with respect to (w.r.t.) $\zeta_k[n]$. To handle this problem, we utilize the liner upper bound, i.e., the first-order Taylor expansion of $\eta_k[n]$, to replace itself. The liner upper bound of $\eta_k[n]$ in the $(t+1)$-th inner loop iteration can be expressed as
\begin{align}
\eta_k[n]\leq& \widehat{\eta}_k[n]\big(\zeta_k[n],\zeta_k^{(t)}[n]\big)=\zeta_k[n]-\big((\zeta_k^{(t)}[n])^2\notag\\
   & +2\zeta_k^{(t)}[n](\zeta_k[n]-\zeta_k^{(t)}[n])\big).
\end{align}

Thus, in the $(t+1)$-th inner loop iteration of the $(l+1)$-th iteration of the Dinkelbach's algorithm, the optimization problem \eqref{eq_opti_resource_association} can be re-expressed
\begin{subequations}\label{eq_opti_resource_association1}
\begin{align}
		&\max _{\boldsymbol{\Xi}_1}~L_\mathrm{tol}(\boldsymbol{\Xi}_1)-\psi^{(l)} E_\mathrm{tol}(\boldsymbol{\Xi}_1)\notag\\
		&\quad~~~-\hat{\rho}\sum\limits_{n=1}^N\sum\limits_{k=1}^K\widehat{\eta}_k[n]\left(\zeta_k[n], \zeta_k^{(t)}[n]\right),\notag\\
		&~\text { s.t. }~\eqref{eq_orig_opti_1}, \eqref{eq_orig_opti_5}, \eqref{eq_users_scheduling}\label{eq_opti_resource_association1_1},
	\end{align}
\end{subequations}
where $\hat{\rho}>0$ denotes the introduced penalty coefficient. The problem \eqref{eq_opti_resource_association1} is a convex optimization problem and can be directly solved by the existing tool such as CVX.
\begin{center}
	\begin{tabular}{p{8.5cm}}
		\toprule[2pt]
		\textbf{Algorithm 1:} The Proposed Iterative Algorithm for Solving the Sub-problem \eqref{eq_opti_resource_association}\\
		\midrule[1pt]
		1: Initialize the feasible point $\big(\mathbf{L}^{(l,0)}, \boldsymbol{\zeta}^{(l, 0)}\big)$; Define the \\\quad tolerance accuracy thresholds as $\widehat{\varepsilon}$; Set the iteration index \\\quad$t$ = 0;  Initialize $\hat{\rho}^{(0)}$\\
		2: \textbf{While} $\widehat{v}>\widehat{\varepsilon}$ or $t=0$ \textbf{do}\\
		3: \quad Solve the optimization problem \eqref{eq_opti_resource_association1} with the given\\\quad~~~$ \boldsymbol{\zeta}^{(l, t)}$ and update $\big(\mathbf{L}^{(l, t+1)}, \boldsymbol{\zeta}^{(l, t+1)}\big)$ with the obtained \\\quad~~~solutions.\\
		4: \quad Calculate $\widehat{v}=\max\limits_{k\in\mathcal{K}, n\in\mathcal{N}}\eta_k[n]$ based on the acquired\\\qquad solutions; Update the penalty coefficients $\hat{\rho}=\omega \hat{\rho}$;\\ \quad~~~Let $t=t+1$. \\
		5: \textbf{end while}\\
		6: Update $\big(\mathbf{L}^{(l+1, 0)}, \boldsymbol{\zeta}^{(l+1, 0)}\big)$ with $\big(\mathbf{L}^{(l, t)}, \boldsymbol{\zeta}^{(l, t)}\big)$. \\
		\bottomrule[2pt]
	\end{tabular}
\end{center}

To solve problem \eqref{eq_opti_resource_association}, we propose an iterative algorithm which is summarized as Algorithm 1. The primary objective aims to ensure that the binary constraint is met by progressively increasing the penalty coefficient with $\hat{\rho}=\omega\hat{\rho}$, where $\omega >1$ is the scaling factor.

\subsubsection{Designing $\boldsymbol{\Xi}_2$ with given $\boldsymbol{\zeta}$ and $\mathbf{Q}$} After achieving the user scheduling  $\boldsymbol{\zeta}$, we focus on designing the passive beamforming variables with the given UAV's trajectory. In particular, the corresponding optimization problem for passive beamforming can be expressed as
\begin{subequations}\label{eq_opti_passive}
	\begin{align}
		&\max _{\boldsymbol{\Xi}_2}~~ L_\mathrm{tol}(\boldsymbol{\Xi}_2)-\psi^{(l)} E_\mathrm{tol}(\boldsymbol{\Xi}_2),\notag \\
		&\quad~\text { s.t. }\eqref{eq_orig_opti_1}, \eqref{eq_orig_opti_5}-\eqref{eq_orig_opti_8}. \label{eq_opti_passive_1}
	\end{align}
\end{subequations}
Note that, problem \eqref{eq_opti_passive} is a non-convex optimization problem because of the non-convexity of constraints \eqref{eq_off_ua},  \eqref{eq_off_BS} and \eqref{eq_orig_opti_6}. However, we can derive the close-form expression of the optimal reflected phases according to the Theorem \ref{theorem 1}.
\begin{theorem}\label{theorem 1}
	The obtained optimal reflection phases at the $n$-th time slot can be given by
	\begin{align}
		\boldsymbol{\phi}_\mathrm{r}^\mathrm{opt}[n]=&\operatorname{arg}(\operatorname{norm}(\mathbf{h}_\mathrm{rb}^*[n]\circ\mathbf{h}_{\mathrm{r}k}[n])),
	\end{align}
	where the selection of the index $k$ is determined by the condition of $\zeta_k[n]=1$.
	\begin{proof}
The proof of Theorem \ref{theorem 1} is given by Appendix \ref{Apd_A}
	\end{proof}
\end{theorem}

Next, we will focus on designing the reflection amplitudes and transmitted coefficient of STAR-RIS's passive beamforming, i.e., $\{\beta^{m}_\mathrm{r}[n]\}_{m\in\mathcal{M}, n\in \mathcal{N}}$ and $\{\boldsymbol{\Theta}_\mathrm{t}[n]\}_{n\in \mathcal{N}}$. For the non-convex constraints \eqref{eq_off_ua} and \eqref{eq_off_BS}, we first rewrite $\boldsymbol{\vartheta}_\mathrm{r}[n]=\boldsymbol{\Phi}_\mathrm{r}[n]\boldsymbol{\beta}_\mathrm{r}[n]$, where  $\boldsymbol{\vartheta}_\mathrm{r}[n]=\operatorname{diag}(\boldsymbol{\Theta}_\mathrm{r}[n])$, $\boldsymbol{\beta}_\mathrm{r}[n]=[\beta_\mathrm{r}^1[n], \cdots, \beta_\mathrm{r}^m[n], \cdots, \beta_\mathrm{r}^M[n]]^T$, and $\boldsymbol{\Phi}_\mathrm{r}[n]=\operatorname{Diag}(e^{j\boldsymbol{\phi}_\mathrm{r}^\mathrm{opt}[n]})$. Therefore, $R^\mathrm{BS}_k[n]$ and $R^\mathrm{ua}_k[n]$ can be further re-expressed as \begin{align}
	R^\mathrm{BS}_k[n]=&\zeta_k[n]B\log_2\Big(1+\boldsymbol{\beta}_\mathrm{r}^T[n]\mathbf{F}_k[n]\boldsymbol{\beta}_\mathrm{r}[n]\Big),\\
	R^\mathrm{ua}_k[n]=&\zeta_k[n]B\log_2\Big(1+\boldsymbol{\vartheta}_\mathrm{t}^H[n]\mathbf{E}_k[n]\boldsymbol{\vartheta}_\mathrm{t}[n]\Big),
\end{align}
where
\begin{itemize}
\item$\boldsymbol{\vartheta}_\mathrm{t}[n]=\operatorname{diag}(\boldsymbol{\Theta}_\mathrm{t}[n])=\{\beta^1_\mathrm{t}[n]e^{j\phi^1_\mathrm{t}[n]}, \cdots, \\\beta^m_\mathrm{t}[n]e^{j\phi^m_\mathrm{t}[n]}, \cdots, \beta^M_\mathrm{t}[n]e^{j\phi^M_\mathrm{t}[n]}\}$,
\vspace{2mm}
\item$\mathbf{F}_k[n]=\frac{\boldsymbol{\Phi}_\mathrm{r}^H[n]}{\sigma_\mathrm{BS}^2}(\mathbf{h}_\mathrm{BR}^*[n]\circ\mathbf{h}_{\mathrm{r}k}[n])(\mathbf{h}_\mathrm{BR}^*[n]\circ\mathbf{h}_{\mathrm{r}k}[n])^H\boldsymbol{\Phi}_\mathrm{r}$,
\vspace{2mm}
\item $\mathbf{E}_k[n]=(\mathbf{h}_\mathrm{ru}^*\circ\mathbf{h}_{\mathrm{r}k}[n])(\mathbf{h}_\mathrm{ru}^*\circ\mathbf{h}_{\mathrm{r}_k}[n])^H$.
\vspace{2mm}
\end{itemize}

Then, we introduce auxiliary variables $\gamma^\mathrm{BS}_k[n]$ and $\gamma^\mathrm{ua}_k[n]$ which satisfy $\gamma^\mathrm{BS}_k[n]\leq\boldsymbol{\beta}_\mathrm{r}^T[n]\mathbf{F}_k[n]\boldsymbol{\beta}_\mathrm{r}[n]$ and $\gamma^\mathrm{ua}_k[n]\leq\boldsymbol{\vartheta}_\mathrm{t}^H[n]\mathbf{E}_k[n]\boldsymbol{\vartheta}_\mathrm{t}[n]$. Thus, the problem designing reflection amplitudes and transmission coefficients in the ($l+1$)-th iteration of the Dinkelbach's algorithm can be equivalently transformed as
\begin{subequations}\label{eq_opti_passive_trans}
	\begin{align}
		&\hspace{-3mm}~\max _{\mathbf{L} ,\boldsymbol{\beta}_\mathrm{r}, \boldsymbol{\vartheta}_\mathrm{t},\boldsymbol{\gamma} }L_\mathrm{tol}(\boldsymbol{\beta}_\mathrm{r}, \boldsymbol{\vartheta}_\mathrm{t},\boldsymbol{\gamma})-\psi^{(l)}E_\mathrm{tol}(\boldsymbol{\beta}_\mathrm{r}, \boldsymbol{\vartheta}_\mathrm{t},\boldsymbol{\gamma})\notag \\
		&\hspace{-3mm}~\text { s.t. }\eqref{eq_QoS}-\eqref{eq_information_causality_const_BS}, \eqref{eq_orig_opti_5}, \label{eq_opti_passive_trans_1}\\
		&\hspace{-3mm}~~\delta_\mathrm{t}[n]\zeta_k[n]\log_2(1+\gamma^\mathrm{ua}_k[n])\geq l^\mathrm{ua}_k[n],~ \forall k\in\mathcal{K}, n\in\mathcal{N}, \label{eq_opti_passive_trans_2}\\
		&\hspace{-3mm}~~\delta_\mathrm{t}[n]\zeta_k[n]\log_2(1+\gamma^\mathrm{BS}_k[n])\geq l^\mathrm{BS}_k[n],\forall k\in\mathcal{K},n\in\mathcal{N}, \label{eq_opti_passive_trans_3}\\
		&\hspace{-3mm}~~\gamma^\mathrm{ua}_k[n]\leq\boldsymbol{\vartheta}_\mathrm{t}^H[n]\mathbf{E}_k[n]\boldsymbol{\vartheta}_\mathrm{t},~\forall k\in\mathcal{K},~ n\in\mathcal{N}, \label{eq_opti_passive_trans_4}\\
		&\hspace{-3mm}~~\gamma^\mathrm{BS}_k[n]\leq\boldsymbol{\beta}_\mathrm{r}^T[n]\mathbf{F}_k[n]\boldsymbol{\beta}_\mathrm{r}[n],~\forall k\in\mathcal{K},~n\in\mathcal{N}, \label{eq_opti_passive_trans_5}\\
		&\hspace{-3mm}~~(\beta_\mathrm{r}^m[n])^2+(\beta_\mathrm{t}^m[n])^2 \leq 1,~\forall m\in\mathcal{M}, ~n\in\mathcal{N}.\label{eq_opti_passive_trans_6}
	\end{align}
\end{subequations}
where $\boldsymbol{\gamma}\triangleq\{\gamma^\mathrm{ua}_k[n], \gamma^\mathrm{BS}_k[n], k\in\mathcal{K}, n\in\mathcal{N}\}$.
Actually, problem \eqref{eq_opti_passive_trans} is still a non-convex optimization problem due to the non-convex constraints \eqref{eq_opti_passive_trans_4} and \eqref{eq_opti_passive_trans_5}. In order to address this issue, we employ a linear lower bound, specifically the first-order Taylor expansion, to approximate the right-hand side of constraints \eqref{eq_opti_passive_trans_4} and \eqref{eq_opti_passive_trans_5} and replace them accordingly.
It is worth noting that the equality sign in constraint \eqref{eq_orig_opti_6} has been substituted with an inequality sign to establish the convex constraint \eqref{eq_opti_passive_trans_6}, transforming \eqref{eq_opti_passive_trans} into a convex optimization problem.
\begin{proposition}\label{pro_1}
	In fact, this replacement of the equal sign does not impact the fulfilment of the equality constraint, because the constraint \eqref{eq_opti_passive_trans_6} is satisfied with strict equality in the optimal solution of problem \eqref{eq_opti_passive_trans}.
	\begin{proof}
	The proof of Proposition \ref{pro_1} is given by Appendix \ref{Apd_B}.
	\end{proof}
\end{proposition}

\vspace{-4mm} \subsubsection{Designing $\boldsymbol{\Xi}_3$ with given $\boldsymbol{\zeta}$ and $\boldsymbol{\Upsilon}$} Next, we will design the trajectory of the UAV with the obtained $\boldsymbol{\zeta}$ and $\boldsymbol{\Upsilon}$. Specifically, the corresponding optimization problem for  UAV trajectory can be expressed as
\begin{subequations}\label{eq_opti_trajectory}
\begin{align}
		&\max _{ \boldsymbol{\Xi}_3}~L_\mathrm{tol}(\boldsymbol{\Xi}_3)-\psi^{(l)} E_\mathrm{tol}(\boldsymbol{\Xi}_3), ,\notag \\
		&\quad\text { s.t. }\eqref{eq_orig_opti_1}- \eqref{eq_orig_opti_3}, \eqref{eq_orig_opti_5}. \label{eq_opti_trajectory_1}
	\end{align}
\end{subequations}
In fact, problem \eqref{eq_opti_trajectory} is  non-convex  because of constraints \eqref{eq_off_ua} and \eqref{eq_off_BS}. In order to handle this non-convex constraints, we first introduce the auxiliary variables $\lambda_k[n]$ and $\tilde{\lambda}[n]$ with %$\lambda_k[n]\geq d^{\alpha_{\mathrm{r}k}}_{\mathrm{r}k}[n]=\left\|\mathbf{q}_\mathrm{ua}[n]-\mathbf{q}_{k}\right\|^{\alpha_{\mathrm{r}k}}$ and $\tilde{\lambda}[n]\geq d^{\alpha_\mathrm{rb}}_\mathrm{rb}[n]=\left\|\mathbf{q}_\mathrm{BS}-\mathbf{q}_\mathrm{ua}[n]\right\|^{\alpha_\mathrm{rb}}$.
\begin{align}
	\lambda_k[n]\geq d^{\alpha_{\mathrm{r}k}}_{\mathrm{r}k}[n]=\left\|\mathbf{q}_\mathrm{ua}[n]-\mathbf{q}_{k}\right\|^{\alpha_{\mathrm{r}k}},\\
	\tilde{\lambda}[n]\geq d^{\alpha_\mathrm{rb}}_\mathrm{rb}[n]=\left\|\mathbf{q}_\mathrm{BS}-\mathbf{q}_\mathrm{ua}[n]\right\|^{\alpha_\mathrm{rb}}.
\end{align}
Hence, we have
\begin{align}
	R_k^\mathrm{ua}[n]&\geq\tilde{R}_k^\mathrm{ua}[n]=\log_2\Big(1+\frac{\rho p\left|\mathbf{h}^H_\mathrm{ru}\boldsymbol{\Theta}_\mathrm{t}^H[n]\widehat{\mathbf{h}}_{\mathrm{u}k}[n]\right|^2}{\lambda_k[n]\sigma_\mathrm{ua}^2}\Big),
\end{align}	
\begin{align}	
	R_k^\mathrm{BS}[n]&\geq\tilde{R}_k^\mathrm{BS}[n]=\notag\\
	&\log_2\Big(1+\frac{\rho^2p_k[n]\left|\widehat{\mathbf{h}}_\mathrm{BR}^H[n]\boldsymbol{\Theta}_\mathrm{r}^H[n]\widehat{\mathbf{h}}_{\mathrm{u}k}[n]\right|^2}{\lambda_k[n]\tilde{\lambda}[n]\sigma_\mathrm{ua}^2}\Big).
\end{align}
Actually, $\tilde{R}_k^\mathrm{ua}[n]$ is a convex function w.r.t. $\lambda_k[n]$ and $\tilde{R}_k^\mathrm{BS}[n]$ is the jointly convex function w.r.t. $\lambda_k[n]$ and $\tilde{\lambda}[n]$, and thus we can apply the first-order Taylor expansion of  $\tilde{R}_k^\mathrm{ua}[n]$ and $\tilde{R}_k^\mathrm{BS}[n]$ at the given point $(\lambda_k^{(l)}[n], \tilde{\lambda}^{(l)}[n])$ in $(l+1)$-th iteration of the Dinkelbach's algorithm to convert constraints \eqref{eq_off_ua} and \eqref{eq_off_BS} as the following convex constraints
\begin{align}
&\zeta_k[n]\delta_\mathrm{t}\widehat{R}_k^\mathrm{ua}[n]\geq l_k^\mathrm{ua}[n], ~ \forall k\in\mathcal{K}, ~n\in\mathcal{N},\label{eq_22i}\\
&\zeta_k[n]\delta_\mathrm{t}\widehat{R}_k^\mathrm{BS}[n]\geq l_k^\mathrm{BS}[n], ~\forall k\in\mathcal{K}, ~n\in\mathcal{N},\label{eq_22j}
\end{align}
where
\begin{itemize}
	\item$\widehat{R}_k^\mathrm{ua}[n]=\log_2\Big(1+\frac{\rho p\left|\mathbf{h}^H_\mathrm{ru}\boldsymbol{\Theta}_\mathrm{t}^H[n]\widehat{\mathbf{h}}_{\mathrm{r}k}[n]\right|^2}{\lambda^{(l)}_k[n]\sigma_\mathrm{ua}^2}\Big)+(\lambda_k[n]-\lambda^{(s)}_k[n])f(\lambda^{(l)}_k[n])$.
\vspace{2mm}	
\item$\widehat{R}_k^\mathrm{BS}[n]=(\lambda_k[n]-\lambda^{(l)}_k[n])\tilde{f}_1(\lambda^{(l)}_k[n], \tilde{\lambda}^{(l)}[n])+(\tilde{\lambda}[n]-\tilde{\lambda}^{(l)}[n])\tilde{f}_2(\lambda^{(l)}_k[n], \tilde{\lambda}^{(l)}[n])+\log_2\big(1+\frac{\rho^2p\left|\widehat{\mathbf{h}}_\mathrm{BR}^H[n]\boldsymbol{\Theta}_\mathrm{r}^H[n]\widehat{\mathbf{h}}_{\mathrm{r}k}[n]\right|^2}{\lambda^{(l)}_k[n]\tilde{\lambda}^{(l)}[n]\sigma_\mathrm{ua}^2}\big)$.
\vspace{2mm}	
\item$f(\lambda^{(l)}_k[n])=\frac{-\rho p\left|\mathbf{h}^H_\mathrm{ru}\boldsymbol{\Theta}_\mathrm{t}^H[n]\widehat{\mathbf{h}}_{\mathrm{r}k}[n]\right|^2}{\ln2\big(\rho p\left|\mathbf{h}^H_\mathrm{ru}\boldsymbol{\Theta}_\mathrm{t}^H[n]\widehat{\mathbf{h}}_{\mathrm{r}k}[n]\right|^2+\lambda^{(l)}_k[n]\sigma_\mathrm{ua}^2\big)\lambda^{(l)}_k[n]}$.
\vspace{2mm}
	\item$\tilde{f}_1(\lambda^{(l)}_k[n],\tilde{\lambda}^{(l)}[n])=\\	\frac{-\rho^2p\left|\widehat{\mathbf{h}}_\mathrm{BR}^H[n]\boldsymbol{\Theta}_\mathrm{r}^H[n]\widehat{\mathbf{h}}_{\mathrm{r}k}[n]\right|^2}{\ln2\lambda^{(l)}_k[n]\big(\rho^2p\left|\widehat{\mathbf{h}}_\mathrm{BR}^H[n]\boldsymbol{\Theta}_\mathrm{r}^H[n]\widehat{\mathbf{h}}_{\mathrm{r}k}[n]\right|^2+\lambda^{(l)}_k[n]\tilde{\lambda}^{(l)}[n]\sigma_\mathrm{ua}^2\big)}$.
\vspace{2mm}	\item$\tilde{f}_2(\lambda^{(l)}_k[n],\tilde{\lambda}^{(l)}[n])=\\
\frac{-\rho^2p\left|\widehat{\mathbf{h}}_\mathrm{BR}^H[n]\boldsymbol{\Theta}_\mathrm{r}^H[n]\widehat{\mathbf{h}}_{\mathrm{r}k}[n]\right|^2}{\ln2\tilde{\lambda}^{(l)}[n]\big(\rho^2p\left|\widehat{\mathbf{h}}_\mathrm{BR}^H[n]\boldsymbol{\Theta}_\mathrm{r}^H[n]\widehat{\mathbf{h}}_{\mathrm{r}k}[n]\right|^2+\lambda^{(l)}_k[n]\tilde{\lambda}^{(l)}[n]\sigma_\mathrm{ua}^2\big)}$.
\vspace{2mm}
\end{itemize}

 Note that, $E_\mathrm{tol}$ is a non-convex function w.r.t. variable $\mathbf{Q}$ due to the non-convexity of $E_\mathrm{ua}^\mathrm{fly}$. To tackle this problem, we further introduce a non-negative auxiliary variable $\widehat{\mu}[n]$ with $\widehat{\mu}^2[n]\geq(1+\frac{v^4[n]}{4v_0^4})^{\frac{1}{2}}-\frac{v^2[n]}{2v_0^2}$ to obtain the upper bound of $E_\mathrm{ua}^\mathrm{fly}$, which is expressed as
 \begin{align}
 	\widehat{E}^\mathrm{fly}_\mathrm{ua}&=\sum_{n=1}^N\delta_\mathrm{t}\Big(P_0\Big(1+\frac{3v^2[n]}{U_\mathrm{tip}^2}\Big)+\frac{1}{2}\mu\psi qAv^3[n]\notag\\
 	&+\widehat{P}_\mathrm{0}\widehat{\mu}[n]\Big).
 \end{align}
Actually, the introduced constraint $\widehat{\mu}^2[n]\geq\big(1+\frac{v^4[n]}{4v_0^4}\big)^{\frac{1}{2}}-\frac{v^2[n]}{2v_0^2}$ is a non-convex constraint. To handle this non-convex constraint, we first equivalently transform it as $	\widehat{\mu}^2[n]+\frac{v^2[n]}{v_0^2}\geq\frac{1}{\widehat{\mu}^2[n]}.$
Note that $\widehat{\mu}^2[n]+\frac{v^2[n]}{v_0^2}$ is a jointly convex function w.r.t. $\widehat{\mu}[n]$ and $v[n]$, so the first-order Taylor expansion can be  utilized to transform this constraint, and thus we have
\begin{align}
	&g(\widehat{\mu}, v[n])=(\widehat{\mu}^{(l)}[n])^2+2\mu^{(l)}[n](\widehat{\mu}[n]-\widehat{\mu}^{(l)}[n])\notag\\
	&+\frac{2}{v_0^2\delta_\mathrm{t}^2}(\mathbf{q}_\mathrm{ua}^{(l)}[n+1]-\mathbf{q}_\mathrm{ua}^{(l)}[n])^T(\mathbf{q}_\mathrm{ua}[n+1]-\mathbf{q}_\mathrm{ua}[n])\notag\\
	&-\frac{\left\|\mathbf{q}_\mathrm{ua}^{(l)}[n+1]-\mathbf{q}_\mathrm{ua}^{(l)}[n]\right\|^2}{v_0^2\delta_\mathrm{t}^2}\geq\frac{1}{\widehat{\mu}^2[n]}. \label{eq_energy_fly_trans}
\end{align}

As a result, the optimization problem \eqref{eq_opti_trajectory} in the $(l+1)$-th iteration of the Dinkelbach's algorithm can be transformed as
\begin{subequations}\label{eq_opti_trajectory_trans}
	\begin{align}
		&\max _{ \boldsymbol{\Xi} }~~~\mathbf{L}(\boldsymbol{\Xi})-\psi^{(l)}\widehat{E}_\mathrm{tol}(\boldsymbol{\Xi}), \notag \\
		&\quad\text { s.t. }\eqref{eq_QoS}-\eqref{eq_information_causality_const_BS}, \eqref{eq_orig_opti_2}, \eqref{eq_orig_opti_3}, \eqref{eq_orig_opti_5}, \eqref{eq_22i}, \eqref{eq_22j}, \eqref{eq_energy_fly_trans},\label{eq_opti_trajectory_trans_1}\\
		&\qquad~~~\lambda_k[n]\geq\left\|\mathbf{q}_\mathrm{ua}[n]-\mathbf{q}_{k}\right\|^{\alpha_{\mathrm{r}k}},\label{eq_opti_trajectory_trans_2}\\
		&\qquad~~~\tilde{\lambda}[n]\geq\left\|\mathbf{q}_\mathrm{BS}-\mathbf{q}_\mathrm{ua}[n]\right\|^{\alpha_{\mathrm{rb}}},\label{eq_opti_trajectory_trans_3}
	\end{align}
\end{subequations}
where $\boldsymbol{\Xi}=\{\boldsymbol{\Xi}_3, \lambda_k[n], \tilde{\lambda}[n], \mu[n]\}$, $\widehat{E}_\mathrm{tol}=\widehat{E}^\mathrm{fly}_\mathrm{ua}+E_\mathrm{ut}+E_\mathrm{ua}^\mathrm{com}$. Consequently, we can leverage the convex optimization solver, e.g., CVX, to address this problem.
\subsection{Proposed Optimization Algorithm Analysis }\label{Algorithm}
The presented iterative algorithm for solving the original optimization problem \eqref{eq_orig_opti} is summarized as Algorithm 2, which is two-tier iterative algorithm designed to tackle the three subproblems explained in Section \ref{sec:S3}. Specifically,  the inner loop is responsible for solving the binary variable, i.e., user association variable, by progressively increase the penalty coefficient, $\widehat{\rho}$.
$v$ represents the objective function value obtained after the transformation through the Dinkelbach's algorithm. Once $v$ falls below a predefined threshold $\varepsilon$, the proposed algorithm will converge.

Then, the computational complexity of the proposed iterative algorithm is analysed. Specifically, the computational complexity mainly comes from solving the divided three subproblems, i.e., \eqref{eq_opti_resource_association}, \eqref{eq_opti_passive} and \eqref{eq_opti_trajectory}. Regarding the first subproblem, we propose an iterative penalty algorithm that incorporates both Dinkelbach's algorithm to effectively address it. It is presumed that the interior point method will be employed to compute the transformed standard convex optimization problem. Consequently, the computational complexity for resolving the first subproblem can be expressed as $\mathcal{O}_1=\mathcal{O}(L_1(5NK)^{3.5})$, where $L_1$ denotes the total  iterations number of the Algorithm 1. The computational complexity of addressing the subproblem \eqref{eq_opti_passive_trans} can be expressed as $\mathcal{O}_2=\mathcal{O}((6NK+2M)^{3.5})$, \textcolor{blue}{which is significantly lower than the computing complexity associated with using the semidefinite relaxation (SDR) method for addressing this subproblem, given by $\mathcal{O}((4NK+2M^2)^{3.5})$.} For the subproblem 3, the SCA technique is adopted to transform the UAV trajectory problem into a convex optimization problem. The the computational complexity can be determined as $\mathcal{O}_3=\mathcal{O}((4NK+2N)^{3.5})$. Therefore, the total computing complexity of the proposed algorithm can be expressed as $\mathcal{O}_\mathrm{tol}=\widehat{L}(\mathcal{O}_1+\mathcal{O}_2+\mathcal{O}_2)$, where $\widehat{L}$ denotes the total iteration number of the proposed algorithm. Based on the overall computational complexity analysis, it is evident that the computing complexity is intricately connected to both the quantity of sub-time slots ($N$) and the number of elements installed at STAR-RIS ($M$).
\begin{center}
	\begin{tabular}{p{8.5cm}}
		\toprule[2pt]
		\textbf{Algorithm 2:}  Proposed Iterative Algorithm to Handle the Optimization Problem \eqref{eq_orig_opti}\\
		\midrule[1pt]
		1: Initialize feasible point $\big(\mathbf{L}^{(0)}, \boldsymbol{\zeta}^{(0)}, \boldsymbol{\Upsilon}^{(0)}, \mathbf{Q}^{(0)}\big)$; Define \\ ~~~~the tolerance accuracy threshold $\varepsilon$; Set the outer iteration\\~~~index $l$ = 0; Calculate $\psi^{(0)}$ with the given initial feasible\\~~  point.\\
		2: \textbf{While} $v>\varepsilon$  or $\l=0$ \textbf{do}                       \\
		3: \quad Solve the sub-problem \eqref{eq_opti_resource_association} by utilizing Algorithm 1 \\\quad\quad with the given $ \boldsymbol{\Upsilon}^{(l)}$ and $\mathbf{Q}^{(l)}$, and update $\mathbf{L}^{(l+1)}$ and \\\quad\quad $\boldsymbol{\zeta}^{(l+1)}$ with the obtained solution.\\
		4: \quad Solve the sub-problem \eqref{eq_opti_passive_trans} with the given $ \boldsymbol{\zeta}^{(l+1)}$ and\\\quad\quad $\mathbf{Q}^{(l)}$, and update $\mathbf{L}^{(l+1)}$ and $\boldsymbol{\Upsilon}^{(l+1)}$ with the obtained\\\quad\quad solution.\\
		5: \quad Solve the sub-problem \eqref{eq_opti_trajectory_trans} with the given $ \boldsymbol{\zeta}^{(l+1)}$ and\\\quad\quad $\boldsymbol{\Upsilon}^{(l+1)}$, and update $\mathbf{L}^{(l+1)}$ and $\mathbf{Q}^{(l+1)}$ with the obtai-\\\quad\quad ned solution.\\
		6:\quad~Calculate $v=L_\mathrm{tol}^{(l+1)}-\psi^{(l)}E_\mathrm{tol}^{(l+1)}$ based on the obtain-\\\quad\quad ed solutions; Update  $\psi^{(l+1)}=\frac{L_\mathrm{tol}^{(l+1)}}{E_\mathrm{tol}^{(l+1)}}$; Let $l=l+1$.\\
		7: \textbf{end while}      \\
		\bottomrule[2pt]
	\end{tabular}
\end{center}

Actually, the convergence of the proposed algorithm can be guaranteed, since the framework of the block coordinate descent (BCD) algorithm is utilized to address the optimization problem \eqref{eq_orig_opti}, which will ensure that each iteration results in a solution that is at least as good as the previous one and thus the objective function is monotonically non-decreasing versus the iteration. Additionally, we will also verify the convergence of the proposed  algorithm through the simulation results presented in Section \ref{sec:S4}. \textcolor{blue}{To evaluate the quality of the solutions obtained from the proposed algorithm, the SDR method is selected as the algorithm for comparison in Section \ref{sec:S4}, which is commonly utilized for addressing optimization challenges in the field of communication. Since the approximation accuracy of solutions obtained through the SDR method in solving various types of optimization problems has been theoretically demonstrated \cite{luo2010semidefinite}. In addition, the authors of \cite{luo2010semidefinite} thoroughly summarize the theoretical approximation accuracy of solving various optimization problems in the field of communication using SDR methodology. Note that the approximation accuracy represents the ratio between the obtained solution by leveraging the SDR and the theoretically optimal solution. Thus, choosing the SDR method as a comparative benchmark algorithm can provide an effective evaluation of the theoretical gap between the achieved solution and the optimal solution.
}

\section{Simulation Results}\label{sec:S4}
To highlight the effectiveness of the proposed STAR-RIS-aided UAV-enabled MEC scheme, we present the numerical simulation results in this section and compare the their with three benchmark schemes, including: \textbf{1) RIS-aided scheme \cite{Xiao23STAR-RIS}:} In this baseline scheme, two adjacent conventional RISs with $\frac{M}{2}$ elements are adopted to replace the STAR-RIS, where one is the reflecting-only RIS and the other one is the transmission-only RIS; \textbf{2) Fixing trajectory scheme:} This scheme focuses on optimizing $\mathbf{L}$, $\boldsymbol{\zeta}$ and $\boldsymbol{\Upsilon}$ with direct UAV trajectory flying from the initial point to the final point at a consistent speed. \textbf{3) Heuristic scheme:} UAV will traverse each user node based on the pre-defined trajectory at a consistent speed. Similarly, variables $\mathbf{L}$, $\boldsymbol{\zeta}$ and $\boldsymbol{\Upsilon}$ will be optimized in this scheme. \textcolor{blue}{ \textbf{4) SDR scheme:} The SDR method is utilized to optimize the passive beamforming variable $\boldsymbol{\Upsilon}$ in this scheme.}
 In addition, the simulation parameters are listed in Table \ref{tab:table1}.
\begin{table}[h!]\
	\renewcommand\arraystretch{1.35}
	\centering
	\caption{Parameters Setting}
	\label{tab:table1}
	\begin{tabular}{|M{4cm}|M{3.8cm}|}
		\hline		
	 \textbf{Parameters}& \textbf{Symbol and Value}\\
		\hline
			Altitude of UAV&$H=30$ m   \\
		\hline
		Bandwidth & $B=1$ MHz\\
		\hline
		Carrier frequency & $\lambda_\mathrm{c}=2.4$ GHz\\
		\hline
			Effective capacitance coefficient of MEC servers& $\iota_\mathrm{ua}=\iota_\mathrm{BS}=10^{-27}$  \\
		\hline
		Initial/final point of UAV trajectory & $\mathbf{q}_\mathrm{I}=[-40, 0, 30]^T$ m, $\mathbf{q}_\mathrm{F}=[40, 0, 30]^T$ m \\
		\hline
			Maximum flight velocity and acceleration of UAV & $v_{\max}=30$ m/s, $a_{\max}=20$ m/s$^2$\\
		\hline
			Maximum CPU frequency & $F_\mathrm{BS}=20$ GHz  \\
		\hline
		Noise power&$\sigma^2_\mathrm{ua}=\sigma^2_\mathrm{BS}=-100$ dBm  \\
		\hline
		Pass loss exponent&$\alpha_{\mathrm{r}k}=2.3$, $\alpha_{\mathrm{rb}}=2.3$ \\
		\hline
		Scaling factor & $\omega=10$\\
       \hline
	   The number of CPU cycles needed for processing 1-bit of data & $\varrho_\mathrm{ua}=\varrho_\mathrm{BS}=10^3$ cycles/bit  \\
		\hline
	    Time slot  & $\delta_\mathrm{t}=0.2$ s \\
		\hline
		Transmitting power at users & $p=20$ dBm\\
       \hline
     Tolerance accuracy thresholds & $\varepsilon=\varepsilon_1=\varepsilon_2=10^{-3}$\\
      \hline	
	\end{tabular}
\end{table}
\begin{figure}[ht]
	\centering
	\includegraphics[scale=0.45]{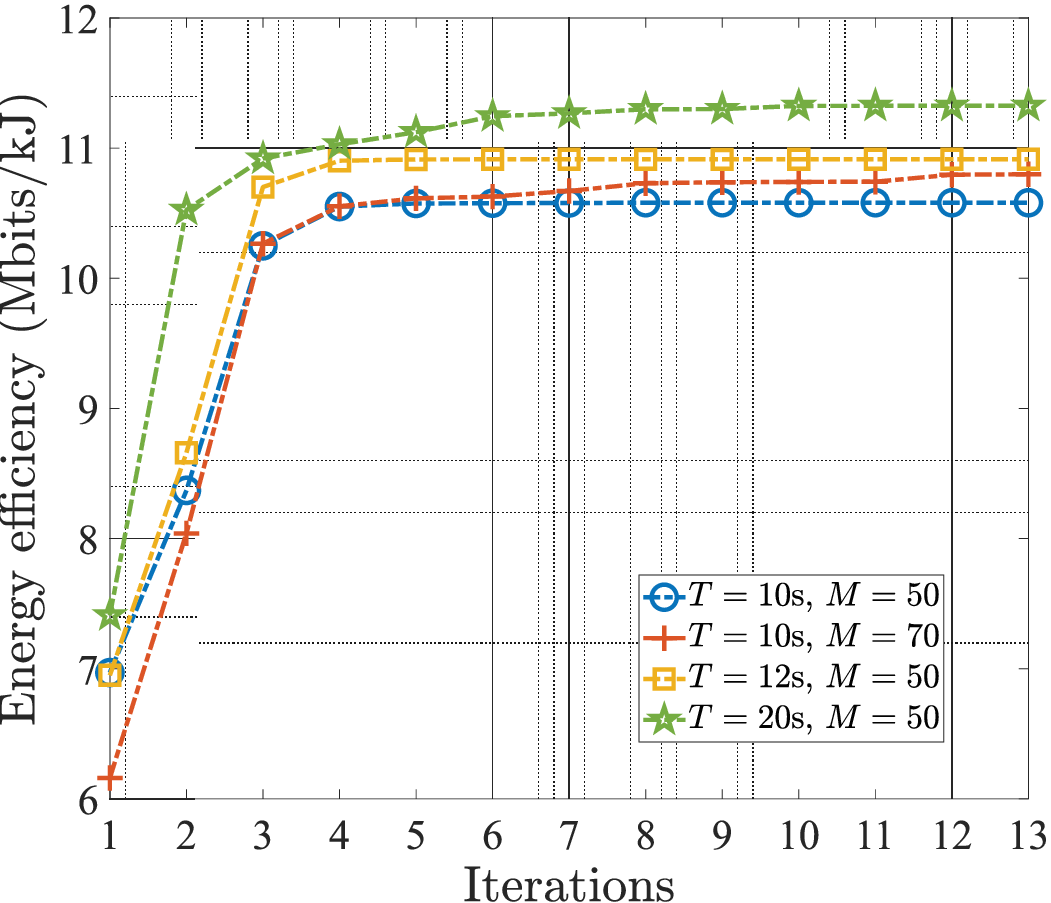}\\
	\caption{Energy efficiency versus iterations index with $L_k=20$ Mbits and $F_\mathrm{ua}=12$ GHz, with different mission period $T$ and the number of elements at STAR-RIS $M$.}\label{fig:iteration}
\end{figure}

In order to assess the convergence of the proposed algorithm, we examine the performance of energy efficiency w.r.t. the number of iteration, which is shown in Fig.  2, taking into account of various mission periods $T$ and the STAR-RIS's number of elements $M$. The outcomes of this investigation are illustrated in Figure \ref{fig:iteration}. Specifically, it is observed that the energy efficiency monotonically increasing with the iteration index and ultimately converges to a specific value. In terms of the four cases under consideration, it is noteworthy that the objective function consistently attains a stable value within a relatively short span of 5-6 iterations.
\begin{figure}[ht]
	\centering
	\includegraphics[scale=0.46]{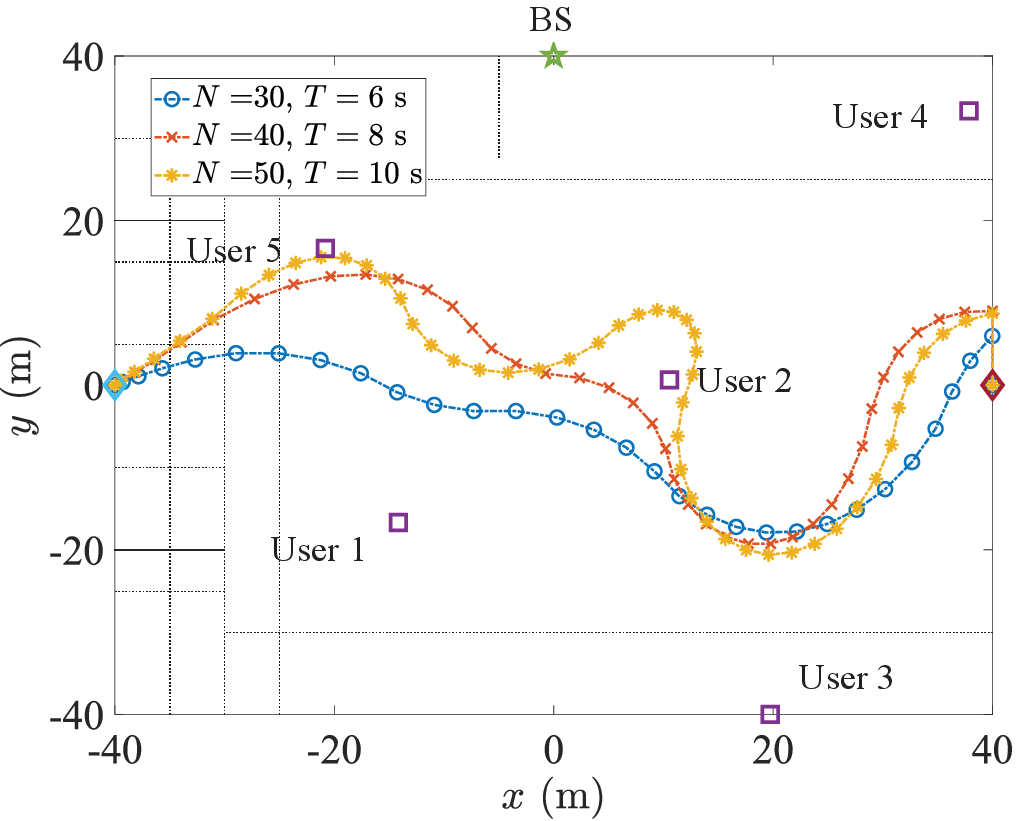}\\
	\caption{UAV trajectory with $M=36$, $L_k=20$ Mbits, $\forall k\in\mathcal{K}$ and $F_\mathrm{ua}= 12$ GHz,  as well as the different  mission period $T$ and $N$.}\label{fig:trajectory_T}
\end{figure}
	
\begin{figure}[ht]
	\centering
	\includegraphics[scale=0.46]{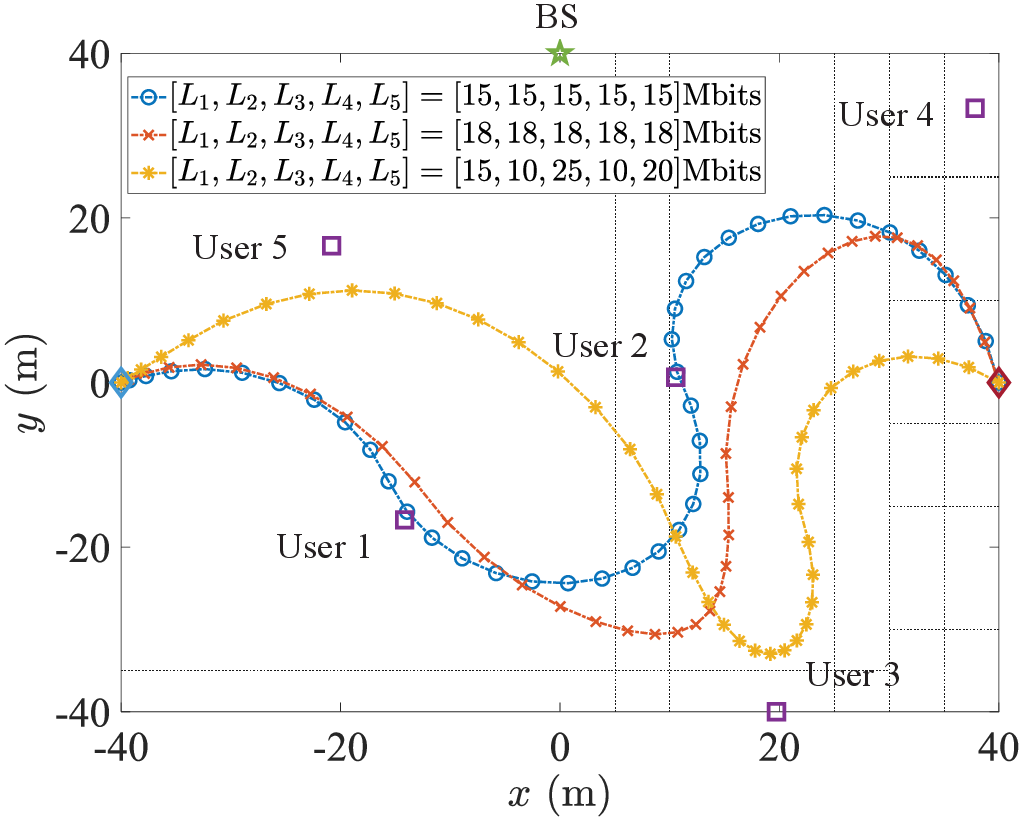}\\
	\caption{UAV trajectory with different $L_k$, with $T=10$ s, $N=50$, $F_\mathrm{ua}= 12$ GHz and $M=36$.}\label{fig:trajectory_L}
\end{figure}

In Fig. \ref{fig:trajectory_T} and Fig. \ref{fig:trajectory_L}, we give the UAV trajectories considering different parameter settings, where $M$  and $F_\mathrm{ua}$ are fixed as $36$ and $12$ GHz, respectively. Specifically, UAV trajectories under different mission period $T$ and time slot number $N$ are shown in Fig. \ref{fig:trajectory_T}, where all the users with computation requirements $L_k=20 $Mbits. We can find that the UAV consistently flies towards User 3, who is the furthest user away from the BS, in order to improve the channel quality between User 3 and the BS under different $T$. This is done to ensure that User 3's minimum task requirement is met. Additionally, it demonstrates that the UAV tends to approach the BS as $T$ increases, which aims to offload larger tasks to the BS, ultimately enhancing the energy efficiency. In Fig. \ref{fig:trajectory_L}, the UAV trajectories is plotted with different $\{L_k\}_{k\in\mathcal{K}}$. It is observed that when all users share the same minimum task requirement, the UAV prioritizes traversing users located at longer distances from the BS, such as User 1, User 2, and User 3. However, if certain users, like User 3 and User 5, require a higher number of offloading task bits, the UAV will approach these users to improve their channel quality and fulfil their task requirements.

\begin{figure}[ht]
	\centering
	\includegraphics[scale=0.45]{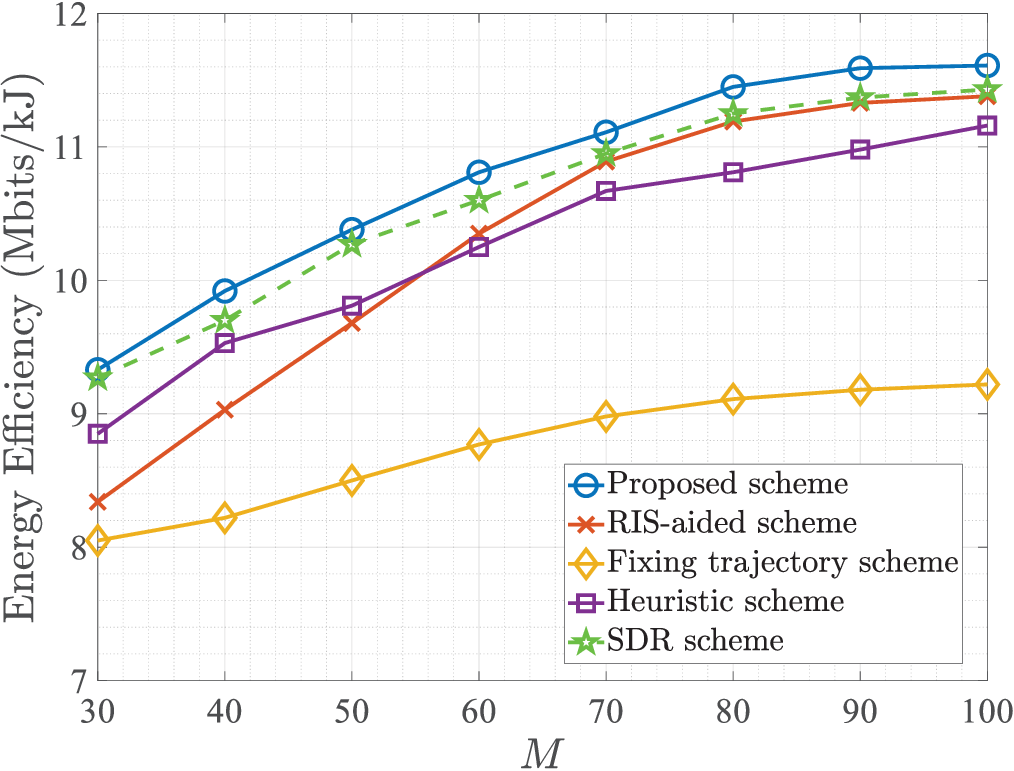}\\
	\caption{Energy efficiency versus the number of elements at STAR-RIS with $N=50$, $F_\mathrm{ua}= 12$ GHz and $L_k=20$ Mbits, $\forall k\in\mathcal{K}$.}\label{fig:element_vs_energy_efficient}
\end{figure}

Next, we investigate the influence of the number of elements equipped at STAR-RIS, i.e., $M$, on the energy efficiency with $N=50$, $F_\mathrm{ua}= 12$ GHz and $L_k=20$ Mbits, $\forall k\in\mathcal{K}$, as shown in Fig. \ref{fig:element_vs_energy_efficient}. Specifically, it can be observed that the energy efficiency of all the schemes increases as $M$ grows, as the additional elements can offer more flexibility to reconfigure the wireless environment. However, the rates of increase gradually decrease as $M$ continues to grow. The proposed scheme offers a greater performance gain in improving the energy efficiency, especially when the number of elements is limited, compared to the conventional RIS-assisted scheme. Additionally, an interesting observation is obtained between the RIS-aided scheme and the heuristic scheme. Specifically, the heuristic scheme achieves higher energy efficiency than the RIS-aided scheme when the $M$ is small. As $M$ increases, the RIS-aided scheme becomes more energy efficient than the heuristic scheme. This indicates that the trajectory of the UAV, with the help of the RIS, has the potential to overcome performance limitations imposed by other system settings. \textcolor{blue}{The presented results in fixing trajectory scheme further verify the importance of the UAV's trajectory in enhancing the system performance. Note that, the proposed scheme illustrates slight advantages through a comparison with the results obtained by the SDR scheme, showcasing the effectiveness of the proposed algorithm.
}
\begin{figure}[ht]
	\centering
	\includegraphics[scale=0.45]{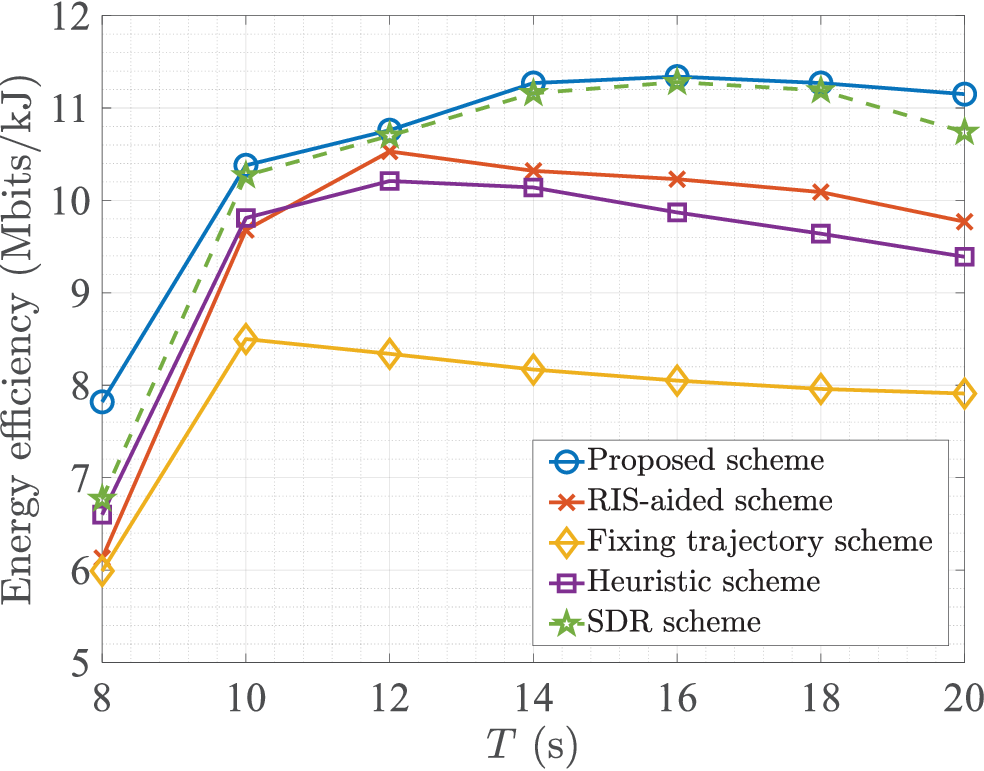}\\
\caption{Energy efficiency versus the mission period $T$ with $M=50$, $F_\mathrm{ua}= 12$ GHz and $L_k=20$ Mbits, $\forall k\in\mathcal{K}$.}\label{fig:ee_t}
\end{figure}

Fig. \ref{fig:ee_t} explores the impact of the mission duration $T$ on energy efficiency when $M$, $F_\mathrm{ua}$ and $L_k, \forall k\in \mathcal{K}$ are respectively set to $50$, $12$ GHz and $20$ Mbits. The results show that the proposed scheme provides the highest performance gain for MEC network, while the scheme without the trajectory optimization exhibits the lowest energy efficiency. Regarding the proposed scheme and the SDR scheme, the energy efficiency consistently increases as $T$ grows from $8$s to $16$s. This can be attributed to the fact that the UAV having more time to optimize its trajectory, thereby improving the channel condition between users/BS and UAV as $T$ increases. However, beyond $16$s, the energy efficiency starts to decline. This is due to the channel quality between users/BS and UAV reaching a saturation point, which may not increase the completed task bits but increase the energy consumption of UAV as $T$ becomes larger. Besides, the conventional RIS-assisted scheme has the similar trend in energy efficiency as the proposed scheme. However, the energy efficiency reaches the limitation at $T=12$s in this scheme, which indicates that STAR-RIS demonstrates a significant potential in achieving a balance between system energy consumption and the volume of the offloaded data when compared to the conventional RIS. In contrast to the suggested scheme, both the heuristic scheme and the scheme with direct UAV trajectory suffer from the earlier performance limitation as $T$ increases. This highlights the significance of the UAV's trajectory optimization in augmenting the energy efficiency of the comprehensive UAV-enabled MEC system. \textcolor{blue}{Furthermore, we can find that the outcomes attained by the suggested algorithm consistently surpass those of the SDR scheme, underscoring the potential and benefits of the proposed algorithm.
}
\begin{figure}[ht]
	\centering
	\includegraphics[scale=0.45]{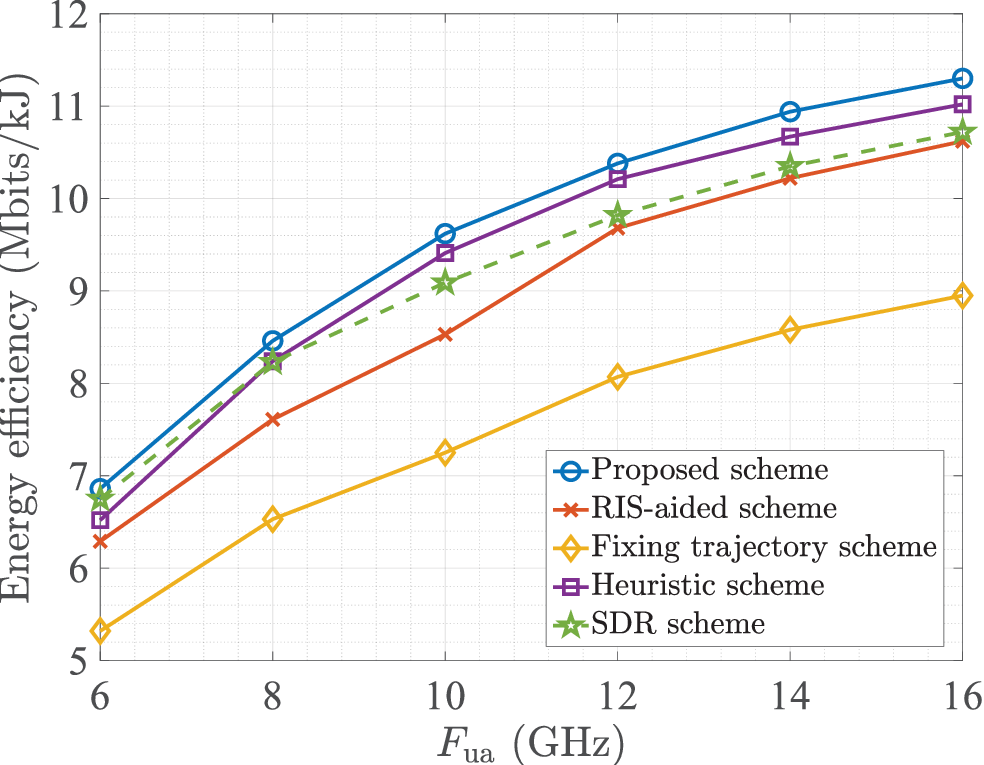}\\
	\caption{Energy efficiency versus the CPU frequency of the MEC server installed at the UAV, $F_\mathrm{ua}$, with $M=50$, $N=50$ and $L_k=10$ Mbits, $\forall k\in\mathcal{K}$.}\label{fig:ee_F_ua}
\end{figure}

Figure \ref{fig:ee_F_ua} illustrates the fluctuating energy efficiency trend corresponding to the CPU frequency of the MEC server deployed on the UAV with parameters $M=50$, $N=50$, and $L_k=10$ Mbits, $\forall k\in\mathcal{K}$. In particular, the escalation of $F_\mathrm{ua}$ correlates with a progressive enhancement in energy efficiency across all scenarios, characterized by a diminishing rate of increase. There are two reasons for this: (\romannumeral 1) By elevating the CPU frequency of the MEC server located at the UAV, users can delegate an expanded array of computational duties to UAV servers for execution. Although this heightened task allocation could potentially elevate energy consumption levels, the pace at which tasks are added outpaces the growth in energy usage, consequently bolstering the system's overall energy efficiency.
(\romannumeral 2) The limitation of achievable offloading rate for users is influenced by system parameters such as allocated task time, number of RIS elements, and users' transmit power. Additionally, the energy consumption of computing tasks assigned to UAVs will increase in a cubic function manner as more tasks are delegated (see \eqref{eq_ua_energy}), contributing to a gradual slowdown in energy efficiency improvement. Furthermore, when compared to the four MEC schemes incorporating trajectory design, i.e., the proposed scheme, RIS-aided scheme, Heuristic scheme and SDR scheme, there is a gradually increase performance gap observed with the scheme lacking UAV trajectory optimization with the growth of $F_\mathrm{ua}$. This indicates that the trajectory optimization plays an important role in balancing the computation bits and energy consumption.

\begin{figure}[ht]
	\centering
	\includegraphics[scale=0.45]{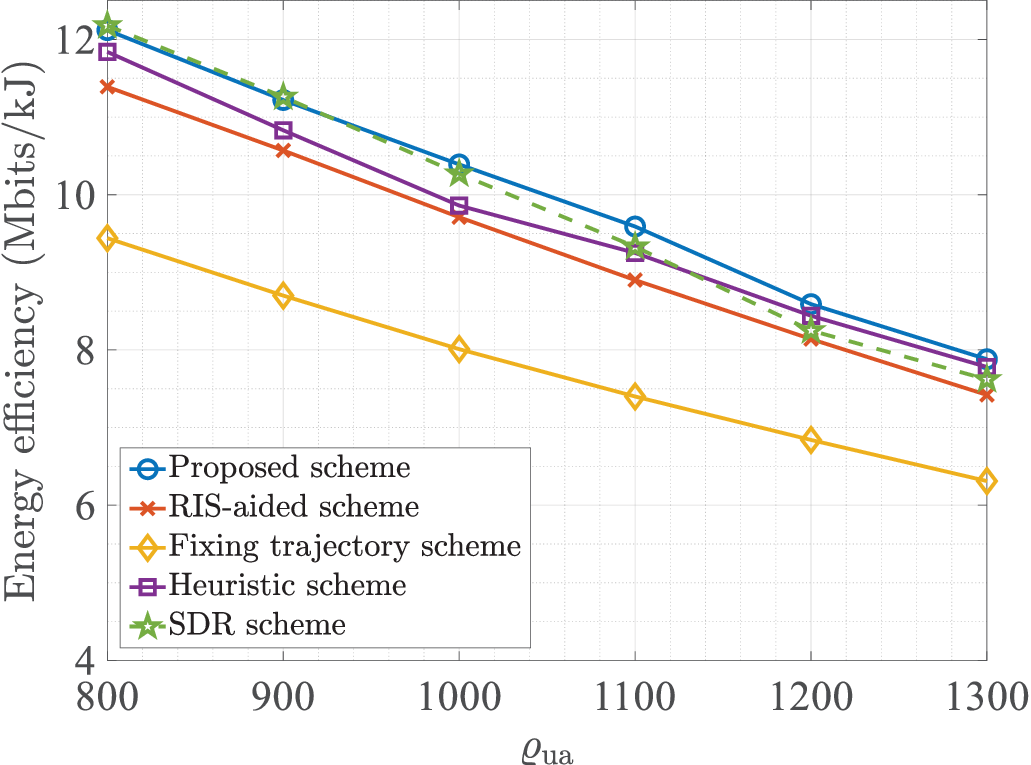}\\
	\caption{Energy efficiency versus CPU cycles required at the UAV for computing 1-bit of task-input data with $M=50$, $N=50$, $F_\mathrm{ua}= 12$ GHz and $L_k=20$ Mbits, $\forall k\in\mathcal{K}$.}\label{fig:ee_varrho}
\end{figure}

The impact of the CPU processing cycles needed by the UAV to compute a single bit of task input data, denoted as $\varrho_\mathrm{ua}$, on energy efficiency is investigated with considerations for $M=50$, $N=50$, and $L_k=20$ Mbits, $\forall k\in\mathcal{K}$. Specifically, it is observed that the energy efficiency declines progressively as the parameter denoted by $\varrho_\mathrm{ua}$ increases for all scenarios. This is due to the fact that the MEC server installed on the UAV must assign additional frequency resources to handle 1-bit data, leading to increased energy consumption for executing offloaded tasks from users. Furthermore, the MEC scheme lacking the optimization of the UAV trajectory demonstrates the least energy efficiency compared with the other four schemes, which highlights the significant potential of optimizing the UAV trajectory design to enhance the energy efficiency of the MEC system. \textcolor{blue}{ It is worth noting that the SDR scheme exhibits the highest performance gain by comparing the obtained simulation results with other four schemes when the $\varrho_\mathrm{ua}\leq 950$. However, the benefit is exceedingly marginal when contrasted with the proposed scheme. The energy efficiency of the SDR scheme diminishes rapidly as the value of $\varrho_\mathrm{ua}$ increases beyond $950$, with the outcomes achieved even falling below those of the Heuristic scheme when $\varrho_\mathrm{ua}$ ranges from $1150$ to $1300$. The results of the simulation demonstrate a consistent and stable reduction when employing the suggested approach, surpassing the performance of the RIS-aided scheme, the fixed trajectory scheme, and the heuristic scheme. This indicates the high robustness of the proposed algorithm.}
\begin{figure}[h]
	\centering
	\begin{subfigure}[t]{0.45\textwidth}
		\centering
		\includegraphics[scale=0.38]{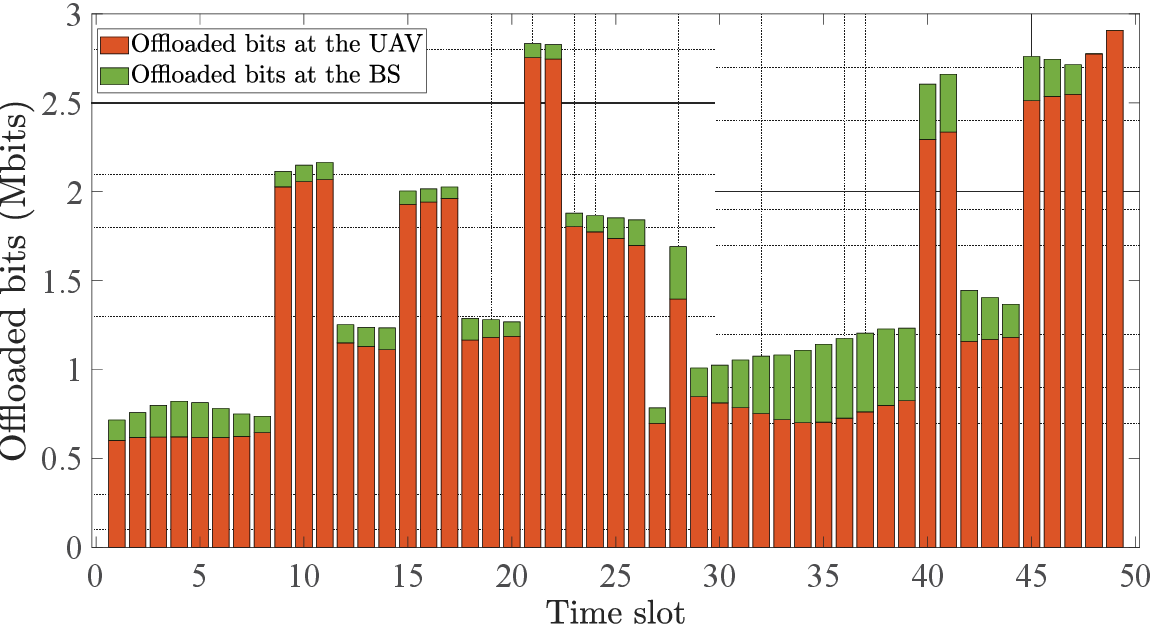}\\
	\caption{The allocation of the offloaded bits at the UAV and the BS versus the time slot.}\label{fig:computation_timeslot}
	\end{subfigure}
	\hfill
	\begin{subfigure}[t]{0.45\textwidth}
		\centering
		\includegraphics[scale=0.38]{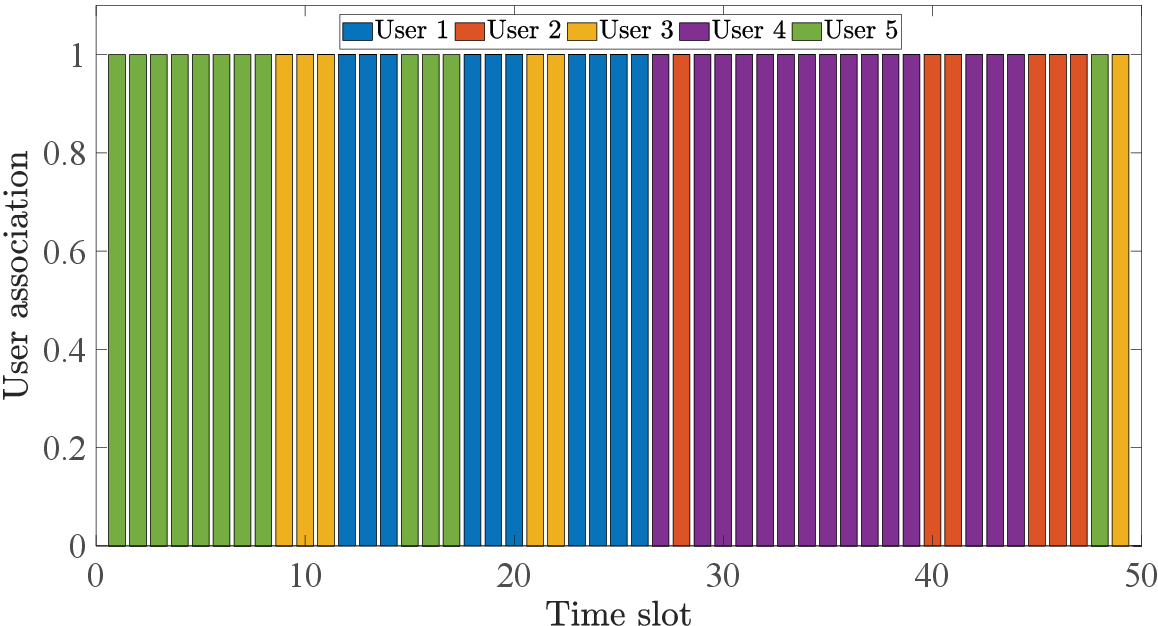}\\
			\caption{User scheduling for task offloading versus the time slot.}\label{fig:user_association_timeslot}
	\end{subfigure}
\caption{The allocation of the offloaded tasks and user scheduling with $M=50$, $F_\mathrm{ua}=6$ GHz, $N=50$ and $L_k=15$ Mbits, $\forall k\in\mathcal{K}$.}\label{fig:timeslot_allocation}
\end{figure}

Finally, to reveal the collaboration between the UAV and the BS for simultaneously processing the offloaded tasks from users, Fig. \ref{fig:timeslot_allocation}(\subref{fig:computation_timeslot}) details the distribution of offloaded bits at the UAV and the BS throughout the mission duration $T$. Specifically, it is noted that the MEC server mounted on the UAV is responsible for processing the majority of computing tasks initiated by users. To meet the demanding computational needs of users, the UAV must primarily handle the processing of offloaded tasks, which is because the UAV's mobility and flexibility are beneficial for enhancing the channel quality between users and the UAV, and the two-path loss of the offloading signals restricts users' ability to offload tasks to the BS. Additionally, we can find that the dynamic trend is evident in the distribution of the offloaded bits at the UAV and the BS, which is corresponding to the UAV's mobility. To show more details, the user scheduling in mission period is presented in Fig. \ref{fig:timeslot_allocation}(\subref{fig:user_association_timeslot}). In particular, it is observed that the considered five users (from User 1 to User 5) are allocated $10$, $6$, $6$, $15$, $12$ time slots to offload their computing tasks to the MEC servers situated at the UAV and the BS, respectively. Note that the final time slot remains unassigned for task offloading, in accordance with the anticipated scheduling. By comparing Fig. \ref{fig:timeslot_allocation}(\subref{fig:computation_timeslot}) and \ref{fig:timeslot_allocation}(\subref{fig:user_association_timeslot}), it is important to highlight that the individuals allocated with more time slots, such as User 4 and User 5, are inclined to reduce the computational bits transferred to the UAV while opting to increase the computational bits directed to the BS within each time slot, which is beneficial for increasing the energy efficiency of the MEC system. To fulfil the minimal task offloading requirement, certain users such as User 2 and User 3, who possess limited time slots, opt to augment the volume of offloaded tasks directed to the MEC server situated on the UAV.
\section{Conclusion}\label{sec:S5}
In this paper, we propose a MEC scheme assisted by STAR-RIS and UAV, which allows the bi-directional task offloading so that users can offload their tasks to the MEC servers situated at the BS and UAV simultaneously. Then, a non-convex optimization problem is established which seeks to maximize the energy efficiency while guaranteeing the QoS constraints for users by jointly designing resources allocation, user scheduling, passive beamforming and the UAV's trajectory. In order to effectively address this non-convex optimization problem, we propose an iterative algorithm that draws inspiration from the Dinkelbach's algorithm and the SCA technique. The proposed iterative algorithm can effectively solve the established problem with guaranteed convergence. The efficacy of the proposed MEC scheme is substantiated through the simulation outcomes in comparison with the several baseline schemes, encompassing the traditional RIS-assisted scheme.
\appendices
\section{Proof of Theorem 1}\label{Apd_A}
Considering that the energy efficiency of the system, denoted as $\frac{L_\mathrm{tol}}{E_\mathrm{tol}}$, excludes the energy utilized for processing user's tasks received at the BS, fully developing the offloaded ability of users to the BS is beneficial for enhancing the overall energy efficiency of the system. When the $k$-th user is selected to offload computing tasks in the $n$-th time slot, i.e., $\zeta_k[n]=1$,  the achievable offloaded rate from the $k$-th user to the BS is given by
\begin{align}
&R^\mathrm{BS}_k[n]=B\log_2\Big(1+\frac{p\left|\mathbf{h}_\mathrm{rb}^H[n]
	\boldsymbol{\Theta}_\mathrm{r}^H[n]\mathbf{h}_{\mathrm{r}k}[n]\right|^2}{\sigma^2_\mathrm{BS}}\Big)\notag\\
&=B\log_2\Big(1+\frac{p\left|\boldsymbol{\varphi}_\mathrm{r}^H[n]\boldsymbol{\Lambda}_\mathrm{r}[n](\mathbf{h}_\mathrm{rb}^*[n]\circ\mathbf{h}_{\mathrm{r}k}[n])
	\right|^2}{\sigma^2_\mathrm{BS}}\Big),
\end{align}
where $\boldsymbol{\varphi}_\mathrm{r}[n]=\big\{e^{j\phi^1_\mathrm{r}[n]}, \cdots, e^{j\phi^m_\mathrm{r}[n]}, \cdots, e^{j\phi^M_\mathrm{r}[n]} \big\}^T$, $\Lambda_\mathrm{r}[n]=\operatorname{Diag}\{\beta^1_\mathrm{r}[n],\cdots, \beta^m_\mathrm{r}[n], \cdots, \beta^M_\mathrm{r}[n]\}$ is a real matrix. To maximize the offloading capacity from the $k$-th user to the BS in the $n$-th time slot, given the reflective amplitudes of the STAR-RIS, i.e., $\boldsymbol{\Lambda}_\mathrm{r}[n]$, the maximum-ratio transmission (MRT) provides the most effective solution for determining $\boldsymbol{\varphi}_\mathrm{r}[n]$ in the $n$-th time slot according to \cite{tse2005fundamentals}. Thus, the optimal $\boldsymbol{\varphi}_\mathrm{r}[n]$, denoted as $\boldsymbol{\varphi}^\mathrm{opt}_\mathrm{r}[n]$, can be derived as
\begin{align}
	\boldsymbol{\varphi}^\mathrm{opt}_\mathrm{r}[n]=\operatorname{norm}(\mathbf{h}_\mathrm{rb}^*[n]\circ\mathbf{h}_{\mathrm{r}k}[n]).
\end{align}
 Therefore, the optimal reflection phases can be expressed as
 \begin{align}
	\boldsymbol{\phi}_\mathrm{r}^\mathrm{opt}[n]=&\operatorname{arg}(\operatorname{norm}(\mathbf{h}_\mathrm{rb}^*[n]\circ\mathbf{h}_{\mathrm{r}k}[n])).
 \end{align}
\section{Proof of Proposition 1}\label{Apd_B}
The contradiction is adopted to prove the Proposition 1. Specifically, we first substitute the optimal reflection phases to the $R_k^\mathrm{BS}[n]$, we have
\begin{align}\label{eq_R_BS}
	R_k^\mathrm{BS}[n]=\log_2\Big(1+\frac{p(\sum_{m=1}^{M}\beta_\mathrm{r}^m[n])^2}{\sigma^2_\mathrm{BS}}\Big).
\end{align}
 It is assumed that the optimal solution of the optimization problem \eqref{eq_opti_passive_trans} does not satisfy the equality in constraint \eqref{eq_opti_passive_trans_6}, i.e., $(\beta_\mathrm{r}^m[n])^2+(\beta_\mathrm{t}^m[n])^2 < 1,~\forall m\in\mathcal{M}, n\in\mathcal{N}$. By fixing the $\beta_\mathrm{t}^m[n]$, we increase the value of $\beta_\mathrm{r}^m[n]$ by employing the scaling factor $\widehat{\omega}>1$ so that  $(\beta_\mathrm{r}^m[n])^2+(\beta_\mathrm{t}^m[n])^2 =1$ remains valid. In this case, we always can achieve a bigger objective function value, since $R_k^\mathrm{BS}[n]$ is a monotonically increasing function w.r.t. $\beta_\mathrm{r}^m[n]$ (see equation \eqref{eq_R_BS}). In other words, scaling up $\beta_\mathrm{r}^m[n]$ can improve the users' task offloading capabilities to the BS, thereby enhancing the overall energy efficiency of the system. The resulting finding conflicts with the initial assumption that $(\beta_\mathrm{r}^m[n])^2+(\beta_\mathrm{t}^m[n])^2 < 1,~ \forall m\in\mathcal{M}, n\in\mathcal{N}$ holds true. Thus, the constraint \eqref{eq_opti_passive_trans_6} is satisfied with strict equality in the optimal solution of problem \eqref{eq_opti_passive_trans}.

\ifCLASSOPTIONcaptionsoff %\emph{a}
  \newpage
\fi
\bibliographystyle{IEEEtran}
\bibliography{MEC-STAR-RIS-STOC}

% Generated by IEEEtran.bst, version: 1.14 (2015/08/26)
\begin{thebibliography}{10}
\providecommand{\url}[1]{#1}
\csname url@samestyle\endcsname
\providecommand{\newblock}{\relax}
\providecommand{\bibinfo}[2]{#2}
\providecommand{\BIBentrySTDinterwordspacing}{\spaceskip=0pt\relax}
\providecommand{\BIBentryALTinterwordstretchfactor}{4}
\providecommand{\BIBentryALTinterwordspacing}{\spaceskip=\fontdimen2\font plus
\BIBentryALTinterwordstretchfactor\fontdimen3\font minus
  \fontdimen4\font\relax}
\providecommand{\BIBforeignlanguage}[2]{{%
\expandafter\ifx\csname l@#1\endcsname\relax
\typeout{** WARNING: IEEEtran.bst: No hyphenation pattern has been}%
\typeout{** loaded for the language `#1'. Using the pattern for}%
\typeout{** the default language instead.}%
\else
\language=\csname l@#1\endcsname
\fi
#2}}
\providecommand{\BIBdecl}{\relax}
\BIBdecl

\bibitem{mao2017survey}
Y.~Mao, C.~You, J.~Zhang, K.~Huang, and K.~B. Letaief, ``{A survey on mobile
  edge computing: The communication perspective},'' \emph{IEEE commun. Surveys
  \& Tuts.}, vol.~19, no.~4, pp. 2322--2358, 2017.

\bibitem{hu2023irs}
X.~Hu, K.-K. Wong, C.~Masouros, and S.~Jin, ``{IRS-Aided Mobile Edge Computing:
  From Optimization to Learning},'' \emph{Intelligent Surfaces Empowered 6G
  Wireless Network}, 2023.

\bibitem{al2015internet}
A.~Al-Fuqaha, M.~Guizani, M.~Mohammadi, M.~Aledhari, and M.~Ayyash, ``{Internet
  of things: A survey on enabling technologies, protocols, and applications},''
  \emph{IEEE Commun. Surveys \& Tuts.}, vol.~17, no.~4, pp. 2347--2376, 2015.

\bibitem{chen2014decentralized}
X.~Chen, ``Decentralized computation offloading game for mobile cloud
  computing,'' \emph{IEEE Trans. Parallel Distrib. Syst.}, vol.~26, no.~4, pp.
  974--983, 2014.

\bibitem{chen2021latency}
C.-L. Chen, C.~G. Brinton, and V.~Aggarwal, ``{Latency minimization for mobile
  edge computing networks},'' \emph{IEEE Trans. Mobile Comput.}, vol.~22,
  no.~4, pp. 2233--2247, 2021.

\bibitem{bi2018computation}
S.~Bi and Y.~J. Zhang, ``Computation rate maximization for wireless powered
  mobile-edge computing with binary computation offloading,'' \emph{IEEE Trans.
  Wireless Commun.}, vol.~17, no.~6, pp. 4177--4190, 2018.

\bibitem{hu2018wireless}
X.~Hu, K.-K. Wong, and K.~Yang, ``Wireless powered cooperation-assisted mobile
  edge computing,'' \emph{IEEE Trans. Wireless Commun.}, vol.~17, no.~4, pp.
  2375--2388, 2018.

\bibitem{shi2021computation}
L.~Shi, Y.~Ye, X.~Chu, and G.~Lu, ``{Computation energy efficiency maximization
  for a NOMA-based WPT-MEC network},'' \emph{IEEE Internet Things J.}, vol.~8,
  no.~13, pp. 10\,731--10\,744, 2021.

\bibitem{jeong2017mobile}
S.~Jeong, O.~Simeone, and J.~Kang, ``{Mobile edge computing via a UAV-mounted
  cloudlet: Optimization of bit allocation and path planning},'' \emph{IEEE
  Trans. Veh. Technol.}, vol.~67, no.~3, pp. 2049--2063, 2017.

\bibitem{gu2021uav}
X.~Gu, G.~Zhang, M.~Wang, W.~Duan, M.~Wen, and P.-H. Ho, ``{UAV-aided
  energy-efficient edge computing networks: Security offloading
  optimization},'' \emph{IEEE Internet Things J.}, vol.~9, no.~6, pp.
  4245--4258, 2021.

\bibitem{xu2021uav}
Y.~Xu, T.~Zhang, Y.~Liu, D.~Yang, L.~Xiao, and M.~Tao, ``{UAV-assisted MEC
  networks with aerial and ground cooperation},'' \emph{IEEE Trans. Wireless
  Commun.}, vol.~20, no.~12, pp. 7712--7727, 2021.

\bibitem{hu2019uav}
X.~Hu, K.-K. Wong, K.~Yang, and Z.~Zheng, ``{UAV-assisted relaying and edge
  computing: Scheduling and trajectory optimization},'' \emph{IEEE Trans.
  Wireless Commun.}, vol.~18, no.~10, pp. 4738--4752, 2019.

\bibitem{hu2020Wireless}
X.~Hu, K.-K. Wong, and Y.~Zhang, ``Wireless-powered edge computing with
  cooperative {UAV}: {T}ask, time scheduling and trajectory design,''
  \emph{IEEE Wireless Commun.}, vol.~19, no.~12, pp. 8083--8098, 2020.

\bibitem{zhang2019joint}
T.~Zhang, Y.~Xu, J.~Loo, D.~Yang, and L.~Xiao, ``{Joint computation and
  communication design for UAV-assisted mobile edge computing in IoT},''
  \emph{IEEE Trans. Ind. Informat.}, vol.~16, no.~8, pp. 5505--5516, 2019.

\bibitem{zhang2019computation}
J.~Zhang, L.~Zhou, F.~Zhou, B.-C. Seet, H.~Zhang, Z.~Cai, and J.~Wei,
  ``{Computation-efficient offloading and trajectory scheduling for multi-UAV
  assisted mobile edge computing},'' \emph{IEEE Trans. Veh. Technol.}, vol.~69,
  no.~2, pp. 2114--2125, 2019.

\bibitem{yang2021ai}
Z.~Yang, M.~Chen, X.~Liu, Y.~Liu, Y.~Chen, S.~Cui, and H.~V. Poor, ``{AI-driven
  UAV-NOMA-MEC in next generation wireless networks},'' \emph{IEEE Wireless
  Commun.}, vol.~28, no.~5, pp. 66--73, 2021.

\bibitem{yang2020offloading}
B.~Yang, X.~Cao, C.~Yuen, and L.~Qian, ``{Offloading optimization in edge
  computing for deep-learning-enabled target tracking by internet of UAVs},''
  \emph{IEEE Internet Things J.}, vol.~8, no.~12, pp. 9878--9893, 2020.

\bibitem{huang2019reconfigurable}
C.~Huang, A.~Zappone, G.~C. Alexandropoulos, M.~Debbah, and C.~Yuen,
  ``Reconfigurable intelligent surfaces for energy efficiency in wireless
  communication,'' \emph{IEEE Trans. Wireless Commun.}, vol.~18, no.~8, pp.
  4157--4170, 2019.

\bibitem{wu2019intelligent}
Q.~Wu and R.~Zhang, ``Intelligent reflecting surface enhanced wireless network
  via joint active and passive beamforming,'' \emph{IEEE Trans. Wireless
  Commun.}, vol.~18, no.~11, pp. 5394--5409, 2019.

\bibitem{yang2023reconfigurable}
B.~Yang, X.~Cao, J.~Xu, C.~Huang, G.~C. Alexandropoulos, L.~Dai, M.~Debbah,
  H.~V. Poor, and C.~Yuen, ``{Reconfigurable intelligent computational
  surfaces: When wave propagation control meets computing},'' \emph{IEEE
  Wireless Commun.}, vol.~30, no.~3, pp. 120--128, 2023.

\bibitem{hu2021reconfigurable}
X.~Hu, C.~Masouros, and K.-K. Wong, ``Reconfigurable intelligent surface aided
  mobile edge computing: From optimization-based to location-only
  learning-based solutions,'' \emph{IEEE Trans. Commun.}, vol.~69, no.~6, pp.
  3709--3725, 2021.

\bibitem{li2021energy}
Z.~Li, M.~Chen, Z.~Yang, J.~Zhao, Y.~Wang, J.~Shi, and C.~Huang, ``{Energy
  efficient reconfigurable intelligent surface enabled mobile edge computing
  networks with NOMA},'' \emph{IEEE Trans Cogn. Commun. Netw.}, vol.~7, no.~2,
  pp. 427--440, 2021.

\bibitem{he2023joint}
W.~He, D.~He, X.~Ma, X.~Chen, Y.~Fang, and W.~Zhang, ``{Joint User Association,
  Resource Allocation, and Beamforming in RIS-Assisted Multi-Server MEC
  Systems},'' \emph{IEEE Trans. Wireless Commun.}, 2023.

\bibitem{qin2023joint}
X.~Qin, Z.~Song, T.~Hou, W.~Yu, J.~Wang, and X.~Sun, ``{Joint Optimization of
  Resource Allocation, Phase Shift and UAV Trajectory for Energy-Efficient
  RIS-Assisted UAV-Enabled MEC Systems},'' \emph{IEEE Trans. Green Commun.
  Netw.}, 2023.

\bibitem{xu2022computation}
Y.~Xu, T.~Zhang, Y.~Liu, D.~Yang, L.~Xiao, and M.~Tao, ``{Computation capacity
  enhancement by joint UAV and RIS design in IoT},'' \emph{IEEE Internet Things
  J.}, vol.~9, no.~20, pp. 20\,590--20\,603, 2022.

\bibitem{zhai2022energy}
Z.~Zhai, X.~Dai, B.~Duo, X.~Wang, and X.~Yuan, ``{Energy-efficient UAV-mounted
  RIS assisted mobile edge computing},'' \emph{IEEE Wireless Commun. Lett.},
  vol.~11, no.~12, pp. 2507--2511, 2022.

\bibitem{duo2023joint}
B.~Duo, M.~He, Q.~Wu, and Z.~Zhang, ``{Joint Dual-UAV Trajectory and RIS Design
  for ARIS-Assisted Aerial Computing in IoT},'' \emph{IEEE Internet Things J.},
  2023.

\bibitem{cao2021reconfigurable}
X.~Cao, B.~Yang, C.~Huang, C.~Yuen, M.~Di~Renzo, D.~Niyato, and Z.~Han,
  ``Reconfigurable intelligent surface-assisted aerial-terrestrial
  communications via multi-task learning,'' \emph{IEEE J. Sel. Areas Commun.},
  vol.~39, no.~10, pp. 3035--3050, 2021.

\bibitem{mu2021simultaneously}
X.~Mu, Y.~Liu, L.~Guo, J.~Lin, and R.~Schober, ``{Simultaneously tansmitting
  and reflecting (STAR) RIS aided wireless communications},'' \emph{IEEE Trans.
  Wireless Commun.}, vol.~21, no.~5, pp. 3083--3098, 2022.

\bibitem{zhang2021intelligent}
S.~Zhang, H.~Zhang, B.~Di, Y.~Tan, M.~Di~Renzo, Z.~Han, H.~V. Poor, and
  L.~Song, ``{Intelligent omni-surfaces: Ubiquitous wireless transmission by
  reflective-refractive metasurfaces},'' \emph{IEEE Trans. Wireless Commun.},
  vol.~21, no.~1, pp. 219--233, 2021.

\bibitem{liu2021star}
Y.~Liu, X.~Mu, J.~Xu, R.~Schober, Y.~Hao, H.~V. Poor, and L.~Hanzo, ``Star:
  Simultaneous transmission and reflection for 360$^\circ$ coverage by
  intelligent surfaces,'' \emph{IEEE Wireless Commun.}, vol.~28, no.~6, pp.
  102--109, 2021.

\bibitem{ahmed2023survey}
M.~Ahmed, A.~Wahid, S.~S. Laique, W.~U. Khan, A.~Ihsan, F.~Xu, S.~Chatzinotas,
  and Z.~Han, ``{A Survey on STAR-RIS: Use Cases, Recent Advances, and Future
  Research Challenges},'' \emph{IEEE Internet Things J.}, 2023.

\bibitem{Xiao23STAR-RIS}
H.~Xiao, X.~Hu, P.~Mu, W.~Wang, T.-X. Zheng, K.-K. Wong, and K.~Yang,
  ``{Simultaneously Transmitting and Reflecting RIS (STAR-RIS) Assisted
  Multi-Antenna Covert Communication: Analysis and Optimization},'' \emph{IEEE
  Trans. Wireless Commun.}, vol.~23, no.~6, pp. 6438--6452, 2024.

\bibitem{xiao2023star}
H.~Xiao, X.~Hu, T.-X. Zheng, and K.-K. Wong, ``{STAR-RIS Assisted Covert
  Communications in NOMA Systems},'' \emph{IEEE Trans. Veh. Technol.}, vol.~73,
  no.~4, pp. 5941--5946, 2024.

\bibitem{xiao2024star}
H.~Xiao, X.~Hu, A.~Li, W.~Wang, Z.~Su, K.-K. Wong, and K.~Yang, ``{STAR-RIS
  Enhanced Joint Physical Layer Security and Covert Communications for
  Multi-antenna mmWave Systems},'' \emph{IEEE Trans. Wireless Commun.}, pp.
  1--1, early access, 2024.

\bibitem{liu2023toward}
Z.~Liu, X.~Li, H.~Ji, H.~Zhang, and V.~C. Leung, ``{Toward STAR-RIS-Empowered
  Integrated Sensing and Communications: Joint Active and Passive Beamforming
  Design},'' \emph{IEEE Trans. Veh. Technol.}, 2023.

\bibitem{xue2024noma}
N.~Xue, X.~Mu, Y.~Liu, and Y.~Chen, ``{NOMA Assisted Full Space
  STAR-RIS-ISAC},'' \emph{IEEE Tran.Wireless Commun.}, 2024.

\bibitem{qin2023joint_}
X.~Qin, Z.~Song, T.~Hou, W.~Yu, J.~Wang, and X.~Sun, ``{Joint Resource
  Allocation and Configuration Design for STAR-RIS-Enhanced Wireless-Powered
  MEC},'' \emph{IEEE Trans. Commun.}, vol.~71, no.~4, pp. 2381--2395, 2023.

\bibitem{zhang2023resource}
Q.~Zhang, Y.~Wang, H.~Li, S.~Hou, and Z.~Song, ``{Resource allocation for
  energy efficient STAR-RIS aided MEC systems},'' \emph{IEEE Wireless Commun.
  Lett.}, vol.~12, no.~4, pp. 610--614, 2023.

\bibitem{aung2024aerial}
P.~S. Aung, L.~X. Nguyen, Y.~K. Tun, Z.~Han, and C.~S. Hong, ``Aerial star-ris
  empowered mec: A drl approach for energy minimization,'' \emph{IEEE Wireless
  Communications Letters}, 2024.

\bibitem{zappone2015energy}
A.~Zappone, E.~Jorswieck \emph{et~al.}, ``Energy efficiency in wireless
  networks via fractional programming theory,'' \emph{Found. Trends Commun.
  Inf. Theory}, vol.~11, no. 3-4, pp. 185--396, 2015.

\bibitem{li2021robust}
S.~Li, B.~Duo, M.~Di~Renzo, M.~Tao, and X.~Yuan, ``{Robust secure UAV
  communications with the aid of reconfigurable intelligent surfaces},''
  \emph{IEEE Trans. Wireless Commun.}, vol.~20, no.~10, pp. 6402--6417, 2021.

\bibitem{mao2023energy}
W.~Mao, K.~Xiong, Y.~Lu, P.~Fan, and Z.~Ding, ``{Energy consumption
  minimization in secure multi-antenna UAV-assisted MEC networks with channel
  uncertainty},'' \emph{IEEE Trans. Wireless Commun.}, 2023.

\bibitem{zhang2022beam}
H.~Zhang, N.~Shlezinger, F.~Guidi, D.~Dardari, M.~F. Imani, and Y.~C. Eldar,
  ``{Beam focusing for near-field multiuser MIMO communications},'' \emph{IEEE
  Trans. Wireless Commun.}, vol.~21, no.~9, pp. 7476--7490, 2022.

\bibitem{zeng2019energy}
Y.~Zeng, J.~Xu, and R.~Zhang, ``{Energy minimization for wireless communication
  with rotary-wing UAV},'' \emph{IEEE Trans. Wireless Commun.}, vol.~18, no.~4,
  pp. 2329--2345, 2019.

\bibitem{luo2010semidefinite}
Z.-Q. Luo, W.-K. Ma, A.~M.-C. So, Y.~Ye, and S.~Zhang, ``Semidefinite
  relaxation of quadratic optimization problems,'' \emph{IEEE Signal Process.
  Mag.}, vol.~27, no.~3, pp. 20--34, 2010.

\bibitem{tse2005fundamentals}
D.~Tse and P.~Viswanath, \emph{Fundamentals of wireless communication}.\hskip
  1em plus 0.5em minus 0.4em\relax Cambridge university press, 2005.

\end{thebibliography}

\end{document}